\documentclass[12pt,a4paper]{article}
\usepackage{epsfig}
\usepackage{amsfonts}
\usepackage{amssymb}
\usepackage{amsmath}
\usepackage{color}
\usepackage{cite}

\usepackage{tikz}
\usetikzlibrary{shapes}
\usetikzlibrary{trees}
\usetikzlibrary{calc}
\usetikzlibrary{decorations.pathmorphing}
\usetikzlibrary{decorations.markings}

\definecolor{grayn}{gray}{0.8}

\begin{document}
\begin{center}

\vspace{1cm}

{\bf \large GRASSMANNIANS AND FORM FACTORS WITH $q^2=0$ IN $\mathcal{N}=4$ SYM THEORY.}
\vspace{2cm}

{\large L. V. Bork$^{1,2}$ A.I. Onishchenko$^{3,4,5}$}\vspace{0.5cm}

{\it $^1$Institute for Theoretical and Experimental Physics, Moscow,
	Russia,\\
	$^2$The Center for Fundamental and Applied Research, All-Russia
	Research Institute of Automatics, Moscow, Russia, \\
	$^3$Bogoliubov Laboratory of Theoretical Physics, Joint
	Institute for Nuclear Research, Dubna, Russia, \\
	$^4$Moscow Institute of Physics and Technology (State University), Dolgoprudny, Russia, \\
	$^5$Skobeltsyn Institute of Nuclear Physics, Moscow State University, Moscow, Russia}\vspace{1cm}

\abstract{In this paper we consider tree level form factors of operators from stress tensor operator supermultiplet with light-like operator   momentum $q^2=0$. We present a conjecture for the Grassmannian integral representation both for these tree level form factors as well as for leading singularities of their loop counterparts.  The presented conjecture was successfully checked by reproducing several known answers in
$\mbox{MHV}$ and $\mbox{N}^{k-2}\mbox{MHV}$, $k\geq3$ sectors together with appropriate soft limits. We also discuss the cancellation of spurious poles and relations between different BCFW representations for such form factors on simple examples.}
\end{center}


\newpage

\tableofcontents{}\vspace{0.5cm}
\renewcommand{\theequation}{\thesection.\arabic{equation}}

\section{Introduction}\label{p1}
In the last years a remarkable progress has been made in
understanding the structure of $S$-matrices (amplitudes) in $\mathcal{N}=4$ SYM and other gauge theories (for a review see \cite{Reviews_Ampl_General,Henrietta_Amplitudes} and reference therein).
This progress became possible due to wide use of new approaches to perturbative computations based on exploration of analytical structure of amplitudes (S-matrix) themselves instead of standard  Feynman diagram computations \cite{Henrietta_Amplitudes}. Introduction of new types of variables (like helicity spinors and momentum twistors) together with superspace formalism  \cite{Nair,DualConfInvForAmplitudesCorch} have also played a key role in the developments made \cite{Henrietta_Amplitudes}.

It was also realized that both tree level amplitudes and leading singularities of loop amplitudes in $\mathcal{N}=4$ SYM allow representation in terms of integrals over Grassmannian manifolds \cite{ArkaniHamed_DualitySMatrix}. This discovery later on led to the
development of the on-shell diagram formalism \cite{Arcani_Hamed_PositiveGrassmannians}  together with exciting ideas concerning geometrical interpretation  of scattering amplitudes in $\mathcal{N}=4$ SYM \cite{ArkaniHamed_UnificationResidues,Arcani_Hamed_PositiveGrassmannians,Hoges_Polytopes,
Masson_Skiner_Grassmaians_Twistors,Arcani_Hamed_Polytopes,Amplituhdron_1,Amplituhdron_2,Amplituhdron_3,Amplituhdron_4,Amplituhdron_5,Amplituhdron_6}. The Grassmannian integral representation is also natural from the point of view of integrability based approaches
\cite{BeisertYangianRev,Staudacher_SpectralReg_New,Beisert_SpectralReg_New,Derkachev_SpectralReg_New,
deleeuwm_2014_1,deleeuwm_2014_2,Frassek_BetheAnsatzYangianInvariants}
to the tree and loop level amplitudes in $\mathcal{N}=4$ SYM.

There is another class of interesting objects in $\mathcal{N}=4$ SYM similar to amplitudes - form factors. Form factors are
 operator matrix elements of the form\footnote{Note that the scattering amplitudes in ''all ingoing" notation may be viewed as form factors of unity operator $\langle p_1^{\lambda_1}, \ldots, p_n^{\lambda_n}|0 \rangle$. }
\begin{equation}
\langle p_1^{\lambda_1}, \ldots,
p_n^{\lambda_n}|\mathcal{O}|0\rangle,
\end{equation}
where $\mathcal{O}$ is some gauge invariant operator which upon acting on the vacuum of the theory produces multi-particle state $\langle p_1^{\lambda_1}, \ldots, p_n^{\lambda_n}|$ with momenta $p_1, \ldots, p_n$ and helicities $\lambda_1, \ldots, \lambda_n$. So, we
can think about form factors as the amplitudes of the processes where
classical current or field, coupled via a gauge invariant
operator $\mathcal{O}$, produces some quantum state $\langle p_1^{\lambda_1}, \ldots,p_n^{\lambda_n}|$.

It is believed, that  $\mathcal{N}=4$ SYM  is very likely to be integrable and the study of form factors within this theory will play the same role as the study of form factors within the context of two dimensional integrable systems (for example, see \cite{FF_in_integrable_sys} and references therein). The form factors should be also useful in better understanding both symmetry properties and structure of the $\mathcal{N}=4$ SYM S-matrix and correlation functions. The direct computations of form factors may help us better understand  the "triality" relations: between amplitudes,
Willson loops and correlation functions (see references in \cite{Henrietta_Amplitudes}) and extra relations for
the amplitudes from \cite{HuotEquation,Twistors_DescentEquation}. Also form factors is an excellent testing laboratory for incorporating
non-planarity and massive (off-shell) states within new on-shell computational methods.

The form factors in $\mathcal{N}=4$ SYM were initially considered in \cite{vanNeerven_InfraredBehaviorFormFactorsN4SYM}, almost 20 years ago. The unique investigation of form factors of non-gauge invariant operators build from single field (off-shell currents) was made in \cite{Perturbiner}. After nearly a decade the investigation of
1/2-BPS form factors was again initiated in \cite{FormFactorMHV_component_Brandhuber,BKV_Form_Factors_N=4SYM}.
Later the form factors of operators from 1/2-BPS and Konishi operator  supermultiplets were intensively investigated both at weak \cite{HarmonyofFF_Brandhuber,BKV_SuperForm,BORK_NMHV_FF,FF_MHV_3_2loop} and strong couplings
\cite{Zhiboedov_Strong_coupling_FF,Strong_coupling_FF_Yang_Gao}.
Attempts to find a geometrical interpretation of form factors of operators from
stress tensor operator supermultiplet were performed in \cite{BORK_POLY}.
More complicated situation of multiple operators were considered in \cite{Roiban_FormFactorsOfMultipleOperators,FormFactorMHV_half_BPS_Brandhuber,FormFactorMHV_Remainder_half_BPS_Brandhuber}. Twistor space based representations of 1/2-BPS and more general  form factors
were considered in \cite{Wilhelm_Twisors_1,Wilhelm_Twisors_2}, see also
\cite{LHC_1,LHC_2,LHC_3} for Lorentz harmonic chiral formulation. Special case of form factors of operators corresponding to "defect insertions" was considered in \cite{BoFeng_BoundaryContributions}.
Integrability properties of 1/2-BPS form factors where investigated in \cite{Wilhelm_Integrability_1,Wilhelm_Integrability_2,Wilhelm_Integrability_3,Wilhelm_Integrability_4,FormFactorsSoftTheorems} and in an important paper \cite{Wilhelm_Grassmannians_Integrability} with an explicit construction based on quantum inverse scattering method. Soft theorems in the context of form factors where considered in \cite{FormFactorsSoftTheorems}. The form factors in theories with maximal supersymmetry in dimensions
different from $D=4$ were investigated in
\cite{FF_ABJM_Young,FF_Sudakov_ABJM_Baianchi,Penati_Santambrogio_ABJM_finite_N,Brandhuber_ABJM_Sudakov2loops}.
Other directions in the study of form factors such as colour-kinematic duality and so on were investigated in
\cite{Henn_Different_Reg_FF,3loopSudakovN4SYM,FF_Colour_Kinematic,masters4loopSudakovN4SYM,Oluf_Tang_Engelund_Lagrangian_Insertion}.

In the present article we are going to consider simplified case of form factors of operators from $\mathcal{N}=4$ SYM stress tensor operator supermultiplet with light-like momentum $q^2=0$ carried by operator. This case is the most simple, yet it captures all essential differences of form factors compared to amplitudes, which are basically originating in different color structure. We  will present a conjecture for Grassmannian representation valid both for these tree level form factors as well as for leading singularities of their loop counterparts.

The study of Grassmannian representations for form factors was initiated in \cite{FormFactorsSoftTheorems,Wilhelm_Grassmannians_Integrability}. The more general case of form factors with $q^2\neq0$ was successfully
considered in \cite{Wilhelm_Grassmannians_Integrability}. In principle one should be able to derive Grassmannian representation for $q^2=0$ case from the results of \cite{Wilhelm_Grassmannians_Integrability} by taking appropriate soft limit with respect to one of two spinor variables parameterizing operator's off-shell momentum $q$. Here, however, we found that it is easier for us to start from scratch and use an approach of \cite{FormFactorsSoftTheorems}. There it was claimed, that Grassmannian integral representation for form factors could be obtained modifying  Grassmannian integral representation for amplitudes and introducing an appropriate regulator of Grassmannian integral with respect to soft limit of operator momentum $q$.

This article is organized as follows. In section 2 we briefly remind the reader the Grassmannian integral representation and on-shell
diagram formalism for amplitudes in $\mathcal{N}=4$ SYM.
In section 3 we introduce the notion of regulated on-shell diagrams as well as discuss possible analogs of top-cell diagrams for form factors.
Section 4 contains our conjecture for Grassmannian integral representation for form factors of operators from stress-tensor operator supermultiplet with $q^2=0$.  In section 5 we verify our conjecture against known results for $\mbox{MHV}_n$,$~\mbox{N}^{k-2}\mbox{MHV}_{k+1}$, $\mbox{NMHV}_{5}$ form factors. We have also checked that our Grassmannian integral representation correctly reproduces soft limit with respect to operator momentum $q$. Section 6 is devoted to the discussion of the choice of integration contour, its relation to different BCFW representations for tree level form factors and cancellation of spurious poles.
In section 7 and 8 we discuss some of the open questions, possible further developments and give brief summary of the results obtained. The appendixes contain details regarding the structure of form factors of operators from $\mathcal{N}=4$ SYM operator supermultiplet together with the details of Grassmannian integral and BCFW recursion computations

\section{Grassmannians, amplitudes and on-shell diagrams}\label{p2}
It is known already for some time \cite{ArkaniHamed_DualitySMatrix} that tree level $\mbox{N}^{k-2}\mbox{MHV}_n$ amplitudes in $\mathcal{N}=4$ SYM can be written in terms of integrals over Grassmannian manifolds $Gr(n,k)$
\begin{eqnarray}\label{GrassmannianIntegralLambda}
 A_n^{(k)}(\{\lambda_i,\tilde{\lambda}_i,\eta_i\})&=&
 \int_{\Gamma} \frac{d^{n\times k}C_{al}}{Vol[GL(k)]}\frac{1}{M_1...M_n}
 \prod_{a=1}^k
 \delta^{2}\left(\sum_{l=1}^n C_{al}\tilde{\lambda}_l\right)
 \delta^{4}\left(\sum_{l=1}^n C_{al}\eta_l\right)\times\nonumber\\
 &\times&\prod_{b=k+1}^n
 \delta^{2}\left(\sum_{l=1}^n \tilde{C}_{bl}\lambda_l\right).
\end{eqnarray}
The points of Grassmannian manifold $Gr(k,n)$ are given by complex $k$-planes in $\mathbb{C}^n$ space passing through its origin. For example, the Grassmannian $Gr(1,2)$ is equivalent to projective complex space $Gr(1,2)=\mathbb{C}\mathbb{P}$.
Each $k$-plane may be parameterized by $k$ $n$-vectors in $\mathbb{C}^n$ or equivalently by $n\times k$ matrix
($C$ matrix in (\ref{GrassmannianIntegralLambda})). The points of Grassmanian are then given by $k\times n$ matrices $C$ modulo $GL(k)$ transformations related to $k$-plane basis choice. This explains
$Vol[GL(k)]$ factor in the integration measure of (\ref{GrassmannianIntegralLambda}). $\tilde{C}_{al}$ is the orthogonal complement of $C$ defined by condition
\begin{eqnarray}
C\tilde{C}^{T}=\sum_{i=1}^nC_{ai}\tilde{C}_{bi}=0.
\end{eqnarray}
The $GL(k)$ gauge fixing could be performed in a number of ways. For example, in $Gr(3,6)$ case
(which corresponds to the $\mbox{NMHV}_6$ amplitude) one can choose $GL(3)$ gauge as
\begin{eqnarray}
    C=\left( \begin{array}{cccccc}
        1 & 0 & 0 & c_{14} & c_{15} & c_{16} \\
        0 & 1 & 0 & c_{24} & c_{25} & c_{26} \\
        0 & 0 & 1 & c_{34} & c_{35} & c_{36}\end{array} \right).
\end{eqnarray}
$M_i$ in (\ref{GrassmannianIntegralLambda}) are consecutive $k\times k$ minors of $C_{al}$ matrix. That is for example $M_1=1$, $M_2=+c_{14}$ and so on. The minors corresponding to columns $i_1,\ldots , i_k$ will be denoted as $(i_1,\ldots , i_k)$. So, for example in our $Gr(3,6)$ case we can write
\begin{eqnarray}
    M_2=(234)=\left( \begin{array}{ccc}
         0 & 0 & c_{14}  \\
         1 & 0 & c_{24}  \\
         0 & 1 & c_{34} \end{array} \right),~
        (126)=\left( \begin{array}{ccc}
        1 & 0 & c_{16} \\
        0 & 1 & c_{26} \\
        0 & 0 & c_{36}\end{array} \right).
\end{eqnarray}
The integral in (\ref{GrassmannianIntegralLambda}) can be viewed as multidimensional complex integral and computed using multidimensional generalization of Cauchy theorem \cite{ArkaniHamed_DualitySMatrix}. In this case the result of integration will depend on the choice of integration contour $\Gamma$. The choice of integration contour is not unique and different possible choices of the contour give different BCFW representations of the same amplitude. It is important to mention that there also exists the choice of integration contour, which will reproduce
 \emph{leading singularities} of $A_n^{(k)(l)}$ loop amplitudes. It was conjectured that this relation should hold to all orders of perturbation theory \cite{ArkaniHamed_DualitySMatrix}.
Also there is connection between Grassmannian integral representation
and correlation functions of vertex operators (amplitudes) in twistor string theory.

\begin{figure}[t]
 \begin{center}
  \epsfxsize=7cm
 \epsffile{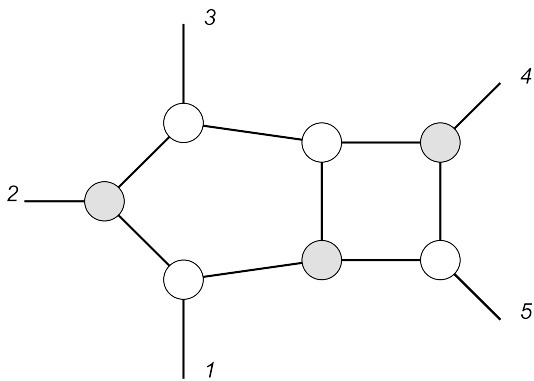}
 \end{center}\vspace{-0.2cm}
 \caption{Top-cell $Gr(2,5)$ on-shell diagram corresponding to $A^{(2)}_5$ amplitude. Grey vertex is $\mbox{MHV}_3$ amplitude, white vertex is $\overline{\mbox{MHV}}_3$ amplitude. }\label{5pointOnSellDiagramMHVAmpl5p}
 \end{figure}
Summing up, the Grassmannian integral representation for the
amplitudes is interesting and useful for the following reasons:
\begin{itemize}
\item It relates different BCFW representation of
tree level amplitudes in $\mathcal{N}=4$ SYM \cite{ArkaniHamed_DualitySMatrix};

\item It could be used to show analytically cancellation of
of spurious poles in BCFW recursion \cite{ArkaniHamed_DualitySMatrix,Arcani_Hamed_PositiveGrassmannians};

\item It gives leading singularities of loop amplitudes in $\mathcal{N}=4$ SYM
\cite{ArkaniHamed_DualitySMatrix,Arcani_Hamed_PositiveGrassmannians};

\item It is claimed \cite{Drummond_Grassmannians_Tduality,Drummond_Yangian_origin_Grassmannian_integral} that the Grassmannian integral representation for amplitudes
(\ref{GrassmannianIntegralLambda}) is the most general form of rational Yangian invariant, which makes all symmetries of the theory manifest. This further points to the integrable structure \cite{Staudacher_SpectralReg_New,Beisert_SpectralReg_New,Derkachev_SpectralReg_New,deleeuwm_2014_1,deleeuwm_2014_2,Frassek_BetheAnsatzYangianInvariants} behind  amplitudes in $\mathcal{N}=4$ SYM (at least at tree level);

\item It relates amplitudes in $\mathcal{N}=4$ SYM and in twistor string theories (see for example \cite{ArkaniHamed_UnificationResidues}).
\end{itemize}

Recently deep insights into the structure of the Grassmannian integral representation of  amplitudes in $\mathcal{N}=4$ SYM were made using so called \emph{on-shell diagram} formalism \cite{Arcani_Hamed_PositiveGrassmannians}. On-shell diagrams are a special type of diagrams build from 3 - point $\mbox{MHV}$ and $\overline{\mbox{MHV}}_3$ vertexes (amplitudes).
$\mbox{MHV}_3$ and $\overline{\mbox{MHV}}_3$ amplitudes themselves can be written in terms of integrals over "small" Grassmannians:
\begin{eqnarray}
 A_3^{(2)}(\{\lambda_i,\tilde{\lambda}_i,\eta_i\})&=&
 \int \frac{d\alpha_{1}}{\alpha_{1}}\frac{d\alpha_{2}}{\alpha_{2}}~
 \delta^{2}\left(\tilde{\lambda}_1+\alpha_1\tilde{\lambda}_3\right)
 \delta^{2}\left(\tilde{\lambda}_2+\alpha_2\tilde{\lambda}_3\right)
 \times\delta^{2}\left(\lambda_3+\alpha_1\lambda_1+\alpha_2\lambda_2\right)\times\nonumber\\
 &\times&\hat{\delta}^{4}\left(\eta_1+\alpha_1\eta_3\right)
 \hat{\delta}^{4}\left(\eta_2+\alpha_2\eta_3\right),
\end{eqnarray}
\begin{eqnarray}
 A_3^{(1)}(\{\lambda_i,\tilde{\lambda}_i,\eta_i\})&=&
 \int \frac{d\beta_{1}}{\beta_{1}}\frac{d\beta_{2}}{\beta_{2}}~
 \delta^{2}\left(\lambda_1+\beta_1\lambda_3\right)
 \delta^{2}\left(\lambda_2+\beta_2\lambda_3\right)
 \times\delta^{2}\left(\tilde{\lambda}_3+\beta_1\tilde{\lambda}_1+\beta_2\tilde{\lambda}_2\right)
 \times\nonumber\\
 &\times&
 \hat{\delta}^{4}\left(\eta_3+\beta_1\eta_1+\beta_2\eta_2\right).
\end{eqnarray}
Gluing $\mbox{MHV}_3$ and $\overline{\mbox{MHV}}_3$ vertexes together with "on-shell propagators" (edges)
\begin{eqnarray}
\int\frac{d^2\lambda_I~d^2\tilde{\lambda}_I~d^4\eta_I}{U(1)}, \label{onshellprop}
\end{eqnarray}
and integrating over internal edge spinor and Grassmann variables in (\ref{onshellprop}) we get integrals over larger Grassmannian submanifolds. See Fig.\ref{5pointOnSellDiagramMHVAmpl5p}
as an example of particular on-shell diagram. So, one can always rewrite a combination of vertexes and edges corresponding to
any given on-shell diagram as an integral over some submanifold of Grassmannian $G(n,k)$
\begin{eqnarray}\label{Omega}
\Omega&=&\int \prod_{i=1}^{n_w}\frac{d\alpha_{1i}}{\alpha_{1i}}\frac{d\alpha_{2i}}{\alpha_{2i}}
 \prod_{j=1}^{n_g}\frac{d\beta_{1i}}{\beta_{1i}}\frac{d\beta_{2i}}{\beta_{2i}}
\prod_{m=1}^{n_I}\frac{1}{U(1)_m}
 \times\nonumber\\
 &\times&\prod_{a=1}^k
 \delta^{2}\left(\sum_{l=1}^n C_{al}[\vec{\alpha}]\tilde{\lambda}_l\right)
 \delta^{4}\left(\sum_{l=1}^n C_{al}[\vec{\alpha}]\eta_l\right)\prod_{b=k+1}^n
 \delta^{2}\left(\sum_{l=1}^n \tilde{C}_{bl}[\vec{\alpha}]\lambda_l\right).
\end{eqnarray}
Here ${\alpha_{1i},\alpha_{2i},\beta_{1i},\beta_{2i}}\equiv\vec{\alpha}$ are edge variables, $n_w$ is the number of white vertexes in on-shell diagram, $n_g$ is the number of gray vertexes and $n_I$ is the number of internal lines. The parameters of Grassmannian $k$ and $n$ are related to the number
of white $n_w$ and gray $n_g$ vertexes together with the number of internal lines $n_I$ of the on-shell diagram as
\begin{eqnarray}
k=2n_g+n_w-n_I,~n=3(n_g+n_w)-n_I.
\end{eqnarray}

Explicit expressions for  $C_{al}[\vec{\alpha}]$ could be found through
the gluing procedure described above, which is highly inefficient however. A more efficient way to express $k\times n$ matrix $C$ in terms of the reduced\footnote{The number of degrees of freedom $d$ of a general on-shell diagram is given by the number of its edges minus number of its vertexes $d = n_I - (n_g + n_w)$ (we subtract $GL(1)$ gauge redundancy associated with every internal vertex)} set of edge variables $\vec{\alpha}$ is by using so called {\it boundary measurement} operation \cite{TotalPositivityGrassmanniansNetworks}. For this purpose one first introduces a {\it perfect matching} $P$, which is a subset of edges in the on-shell diagram, such that every  vertex is the endpoint of exactly one edge in $P$. Next, there is one-to-one correspondence of perfect matching with so called {\it perfect orientation}. A perfect orientation is an assignment of specific orientation to edges, such that each white vertex has a single incoming arrow and each gray vertex has a single outgoing arrow. The edge with a special orientation (directed from gray to white vertex in our case) is precisely the edge belonging to the perfect matching subset \cite{TotalPositivityGrassmanniansNetworks,BipartiteFieldTheories}. Given a perfect orientation all external vertexes are divided into two groups: sources and sinks. Then entries of the matrix $C$ are then given by \cite{TotalPositivityGrassmanniansNetworks}:
\begin{eqnarray}
C_{ij} (\alpha) = \sum_{\Gamma \; \in \; \{i\to j\}} (-1)^{s_{\Gamma}}
\prod_{e \; \in \; \Gamma}\alpha_e^{\{-1,1\}} ,
\end{eqnarray}
where  index $i$ runs over sources, $j$ runs over all external vertexes and $\Gamma$ is an oriented path from $i$  to $j$ consistent with perfect orientation. If the edge is traversed in the direction from white to gray vertex\footnote{It is just a convention for assigning edge variables, which could have been chosen differently.}, then the power of edge variable is $1$, and $-1$ when traversing in opposite direction. The $s_{\Gamma}$ in the formula above is the number of sources strictly between vertexes $i$ and $j$.

One can also think of on-shell diagrams with fixed values of $n$ and $k$ as the integrals over some differential form $d\Omega$ \cite{Arcani_Hamed_PositiveGrassmannians}. In this
sense the general on-shell diagram with fixed values of $n$ and $k$ is the function of integration contour. Next, not all points of Grassmannian in the $\int d\Omega$ integral give nontrivial contributions, but only those belonging to the so called positive Grassmannian $Gr_+(k,n)$ \cite{Arcani_Hamed_PositiveGrassmannians}.
Positive Grassmannian $Gr(k,n)_+$ is a
submanifold in $Gr(k,n)$ defined by the condition that its
points described by $C$ - matrix have strictly positive (cyclically)  consecutive minors. The Grassmanian $Gr(k,n)_+$ could be decomposed into a nested set of submanifolds (called cells) depending on linear dependencies of consecutive column of $C_{al}$ (positroid stratification) \cite{Arcani_Hamed_PositiveGrassmannians}.
The submanifolds (positroid cells) with larger number of linear dependent columns are \emph{the boundaries} of submanifolds with smaller number of linear dependent columns in $C_{al}$. The submanifold
of $Gr(k,n)_+$ containing points, whose coordinates $C_{al}$
contain no linear dependent sets of columns, is called \emph{top-cell}.

\begin{figure}[t]
	\begin{center}
		\epsfxsize=6cm
		\epsffile{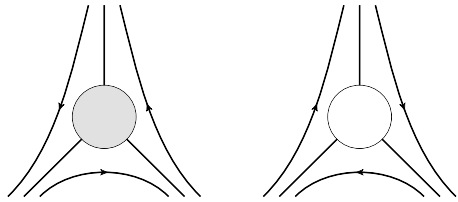}
	\end{center}\vspace{-0.2cm}
	\caption{Elementary permutations ("paths") associated with individual $\mbox{MHV}_3$
		and $\overline{\mbox{MHV}}_3$ vertexes in the on-shell diagrams. }\label{PathesForVertexesAmpl}
\end{figure}

\begin{figure}[t]
	\begin{center}
		\epsfxsize=6cm \epsffile{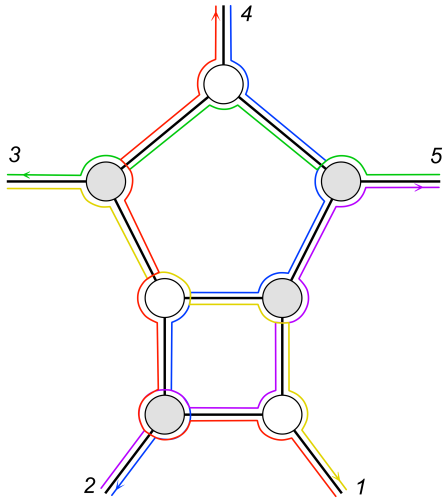}
	\end{center}\vspace{-0.2cm}
	\caption{Example of permutation $\sigma=\{4,5,6,7,8\}$
		associated with $Gr(3,5)$
		on-shell diagram corresponding to $A_5^{(3)}$ amplitude.}
	\label{PathesForNMHV5Ampl}
\end{figure}

There is a correspondence between every submanifold (\emph{positroid cell}) of $Gr(k,n)_+$ mentioned above, \emph{decorated permutation}\footnote{A decorated permutation is an injective map $\sigma:\{1,\ldots,n\}\mapsto\{1,\ldots,2n\}$, such that
$a\leq\sigma(a)\leq a+n$. Taking $\sigma~\mbox{mod}~n$ will give us ordinary permutation. The permutation corresponding to particular on-shell diagram can be obtained by moving along left-right path.
See Figs. \ref{PathesForVertexesAmpl} and \ref{PathesForNMHV5Ampl}.}
and some sub-set\footnote{There are actually equivalent classes
of on-shell diagrams which are labeled by the same permutation. There are also graphical rules (square move and merger/unmerge moves),
which  transform one equivalent diagram into another \cite{Arcani_Hamed_PositiveGrassmannians}.} of all possible on-shell diagrams (the number of faces $F$ of the diagram must be less or equal
to the dimension of $Gr(k,n)_+$ Grassmannian, $dim[Gr(k,n)_+]=k(n-k)$).
Such on-shell diagrams (corresponding integrals $\int d\Omega$)
are given by the rational functions of external kinematical data
$\{\lambda_i,\tilde{\lambda}_i,\eta_i\}$ only. As rational functions on-shell diagrams have poles. These \emph{poles are
in one to one correspondence with the boundaries of cells} in $Gr(k,n)_+$
to which on-shell diagrams correspond to \cite{Arcani_Hamed_PositiveGrassmannians}.

\begin{figure}
	\centering	
	\begin{equation}
	\begin{tikzpicture}[baseline={($(n1.base) - (0,0)$)},transform shape, scale=1]
	\node[right] (n1) at (6,0) {$1$};
	\node[right] (n2) at (6,-0.6) {$2$};
	\node[right] (nk) at (6,-2.4) {$k$};
	\node[right] (nvdots) at (5,-1.6) {$\scalebox{2}{\vdots}$};
	\node[below] (nn) at (0,-3.07) {$n$};
	\node[below] (nnm1) at (1,-3) {$n-1$};
	\node[below] (nkp1) at (4.5,-3) {$k+1$};
	\node[right] (ncdots) at (2.5,-2.8) {$\scalebox{2}{\ldots}$};
	\draw (0,0) -- (6,0);
	\draw (0,-0.6) -- (6,-0.6);
	\draw (0,-1.2) -- (6,-1.2);
	\draw (0,-2.4) -- (6,-2.4);
	\draw (0,0) -- (0,-3);
	\draw (1,0) -- (1,-3);
	\draw (2,0) -- (2,-3);
	\draw (4.5,0) -- (4.5,-3);
	\end{tikzpicture}
	\qquad\qquad
	\begin{tikzpicture}[baseline={($(n1.base) + (0,0.5)$)},transform shape, scale=1]
	\node[right] (n1) at (1.5,-0.7) {$\Rightarrow$};
	\node[right] (n2) at (1.5,-1.8) {$\Rightarrow$};
	\node[right] (n3) at (1.5,-2.9) {$\Rightarrow$};
	\coordinate (nc1) at (3,-0.505) {};
	\coordinate (nc2) at (2.91,-1.8) {};
	\coordinate (nc3) at (2.8,-3);
	\coordinate (nc4) at (3.1,-2.8);
	\draw (0.2,-0.5) -- (0.8,-0.5);
	\draw (0.5,-0.5) -- (0.5,-0.9);
	\draw (2.7,-0.5) -- (3.3,-0.5);
	\draw (3,-0.5) -- (3,-0.9);
	\draw[fill,white] (nc1) circle [radius=0.08];
	\draw[black] (nc1) circle [radius=0.08];
	\draw (0.4,-1.5) -- (0.4,-2.1);
	\draw (0.4,-1.8) -- (0.7,-1.8);
	\draw (2.9,-1.5) -- (2.9,-2.1);
	\draw (2.9,-1.8) -- (3.2,-1.8);
	\draw[fill,grayn] (nc2) circle [radius=0.08];
	\draw[black] (nc2) circle [radius=0.08];
	\draw (0.2,-2.9) -- (0.8,-2.9);
	\draw (0.5,-2.6) -- (0.5,-3.2);
	\draw (nc3) -- (nc4);
	\draw (nc3) -- (2.5,-3);
	\draw (nc3) -- (2.8,-3.3);
	\draw (nc4) -- (3.4,-2.8);
	\draw (nc4) -- (3.1,-2.5);
	\draw[fill,white] (nc3) circle [radius=0.08];
	\draw[black] (nc3) circle [radius=0.08];
	\draw[fill,grayn] (nc4) circle [radius=0.08];
	\draw[black] (nc4) circle [radius=0.08];
	\end{tikzpicture}
	\nonumber
	\end{equation}
	\caption{Top-cell on-shell diagram for $A_n^{(k)}$ on-shell amplitude.}
	\label{TopCellAkn}
\end{figure}
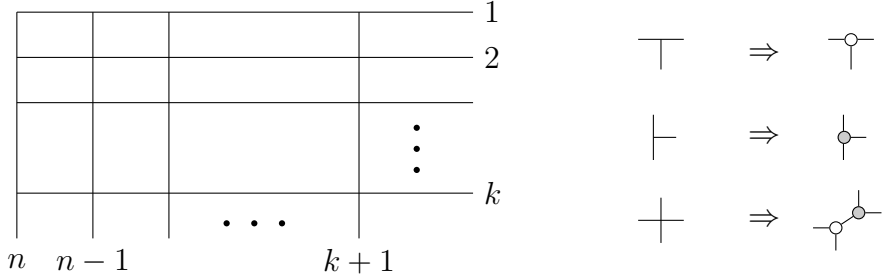

\begin{figure}[t]
\begin{center}
  \epsfxsize=13cm
 \epsffile{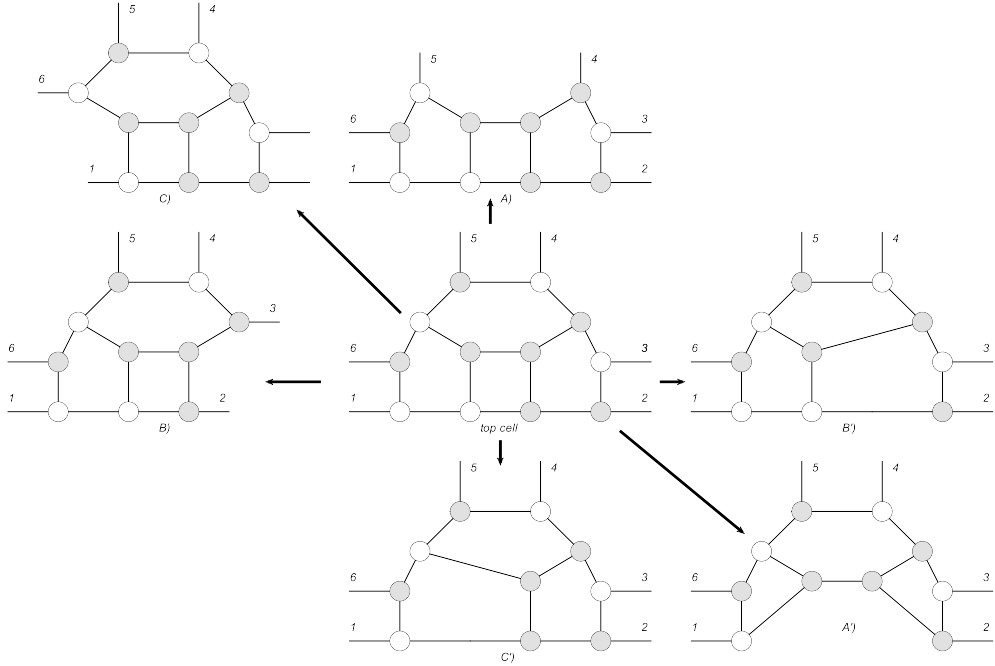}
 \end{center}\vspace{-0.2cm}
 \caption{Top cell $Gr(3,6)$ on-shell diagram and on-shell diagrams
 corresponding to its codimension one boundaries. These on-shell diagrams describe different factorization channels of
 $\mbox{NMHV}_6$ amplitude.
 $A)=A_6^{(2)}R_{136}$, $B)=A_6^{(2)}R_{146}$ and $C)=A_6^{(2)}R_{135}$,
 so that $A)+B)+C)=A_6^{(3)}$ for the $[1,2\rangle$ BCFW shift representation (see appendix \ref{aB}).
 The sum of other three terms gives $[2,3\rangle$
 BCFW shift representation of the same amplitude.}\label{NMHV6AmpltopCell}
 \end{figure}

Within on-shell diagram formalism the BCFW recursion for the tree-level amplitudes $A_n^{(k)}$ is reproduced as follows \cite{Arcani_Hamed_PositiveGrassmannians}. First, one
takes top-cell of $Gr(k,n)_+$ corresponding to a permutation which is a cyclic shift by $k$
\begin{eqnarray}
A_n^{(k)} : \quad\sigma = (k+1, \ldots n,1,\ldots k).
\end{eqnarray}
A representative on-shell top-cell diagram is then constructed as\footnote{See also \cite{HarmonicRmatrices},\cite{Broedel_DictionaryRoperatorsOnshellGraphsYangianAlgebras} for review.} \cite{TotalPositivityGrassmanniansNetworks}: draw  $k$ horizontal lines, $(n-k)$ vertical lines so that the left most and topmost are boundaries and substitute the three and four-crossings as in Fig. \ref{TopCellAkn}. The "boundary" on-shell diagrams corresponding to different  BCFW channels are then obtained by removing $(k-2)(n-k-2)$ edges from top cell diagram (by formal application of the "boundary operator" $\partial$ \cite{Arcani_Hamed_PositiveGrassmannians}). It should be noted that not all edges are removable, but only those which removal lowers the dimension of the on-shell diagram  by exactly one. The exact form of the sum of "boundary" on-shell diagrams can be determined by a formal solution of so called boundary
equation \cite{Arcani_Hamed_PositiveGrassmannians}. See Fig. \ref{NMHV6AmpltopCell} as an example.

It is not hard to show, using particular choice of coordinates on Grassmannian $Gr(k,n)$, that in the case of top-cell diagram the
following identity holds \cite{Arcani_Hamed_PositiveGrassmannians}:
\begin{eqnarray}\label{RelationOfGrassmannianIntegralToTopCell}
\Omega^{top}&=&\int \prod_{i=1}^{n_w}\frac{d\alpha_{1i}}{\alpha_{1i}}\frac{d\alpha_{2i}}{\alpha_{2i}}
 \prod_{j=1}^{n_b}\frac{d\beta_{1i}}{\beta_{1i}}\frac{d\beta_{2i}}{\beta_{2i}}
\prod_{m=1}^{n_I}\frac{1}{U(1)_m}
 \times\nonumber\\
 &\times&\prod_{a=1}^k
 \delta^{2}\left(\sum_{l=1}^n C_{al}[\vec{\alpha}]\tilde{\lambda}_l\right)
 \delta^{4}\left(\sum_{l=1}^n C_{al}[\vec{\alpha}]\eta_l\right)\prod_{b=k+1}^n
 \delta^{2}\left(\sum_{l=1}^n \tilde{C}_{bl}[\vec{\alpha}]\lambda_l\right)=\nonumber\\
&=&\int \frac{d^{n\times k}C_{al}}{Vol[GL(k)]}\frac{1}{M_1...M_n}
 \prod_{a=1}^k
 \delta^{2}\left(\sum_{l=1}^n C_{al}\tilde{\lambda}_l\right)
 \delta^{4}\left(\sum_{l=1}^n C_{al}\eta_l\right)\prod_{b=k+1}^n
 \delta^{2}\left(\sum_{l=1}^n \tilde{C}_{bl}\lambda_l\right).\nonumber\\
\end{eqnarray}
That is, top cell on-shell diagram is given by our initial Grassmannian integral  (\ref{GrassmannianIntegralLambda}). Finally, we would like to note, that the fact that only points of $Gr(k,n)_+$ Grassmannian give nontrivial contribution to Grassmannian integral is closely related to ideas that amplitudes in $\mathcal{N}=4$ SYM may be interpreted as the volume of some geometrical object.


\section{Form factors with $q^2=0$ and regulated on-shell diagrams}\label{p3}

Let us now proceed with the generalization of on-shell diagram formalism and
Grassmannian integral representation for the case of form factors of operators from stress tensor operator supermultiplet at $q^2=0$. For this purpose we are going to use the approach of \cite{FormFactorsSoftTheorems}. It is similar to the approach of \cite{Wilhelm_Grassmannians_Integrability} which was already successfully used to derive Grassmannian integral representation of N$^{k-2}$MHV$_n$ form factors with $q^2 \neq 0$. However, compared to \cite{Wilhelm_Grassmannians_Integrability} in \cite{FormFactorsSoftTheorems} we have only considered some particular examples of form factors and did not supplied the conjecture for general N$^{k-2}$MHV$_n$ form factors. Here we will do that, but for a case of form factors at $q^2 = 0$.
\begin{figure}[t]
 \begin{center}
  \epsfxsize=7cm
 \epsffile{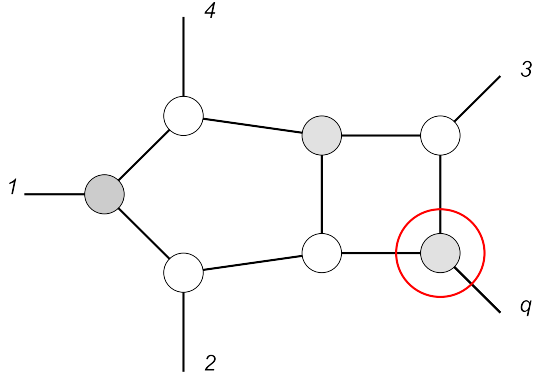}
 \end{center}\vspace{-0.2cm}
 \caption{On-shell diagram with deformed vertex
 $[S(l_1,q,l_2)]^{-1}A_{3}^{(2)}(l_1,l_2,q)$ (red circle). }\label{5pointOnSellDiagramMHVFormF}
 \end{figure}

We begin with the observation that the number of kinematic degrees of freedom
(Weyl spinors associated to momenta of external particles + momentum carried by operator) of n-point super from factors with $q^2=0$ $Z^{(k)}_{n}$ are the same as for
$A^{(k)}_{n+1}$ amplitude. Also note, that MHV form factors of operators from stress tensor operator supermultiplet and MHV amplitudes could be related as:
\begin{eqnarray}
Z^{(2)}_n=S^{-1}(i,q,i+1)A_{n+1}^{(2)}(1,\ldots,i,q,i+1,\ldots,n). \label{MHV_formfactors_amplitudes}
\end{eqnarray}
Here $S^{-1}(i,q,i+1)$ is inverse soft factor which depends on Weyl spinors associated with momenta $p_i,q$ and $p_{i+1}$. This factor could be viewed as some sort of IR regulator. Indeed, the form factor $Z^{(2)}_n$ is regular with respect to the $q\rightarrow0$ limit, while the amplitude $A_n^{(2)}$ is singular. The same will be true also for the general $\mbox{N}^{k-2}\mbox{MHV}$ case. To be more precise, in the  case of tree level amplitudes we have
\begin{eqnarray}
&&A_{n+1}^{(k)}\left(\{\epsilon\lambda_s,\tilde{\lambda}_s,\eta_s\},
{\{\lambda_1,\tilde{\lambda}_1,\eta_1\},...,\{\lambda_n,\tilde{\lambda}_n,\eta_n\}}\right)=
\nonumber\\
&&\left(\frac{\hat{S}_1}{\epsilon^2}+\frac{\hat{S}_2}{\epsilon}\right)
A_{n}^{(k)}\left({\{\lambda_1,\tilde{\lambda}_1,\eta_1\},
...,\{\lambda_n,\tilde{\lambda}_n,\eta_n\}}\right)
+reg.,~\epsilon \rightarrow0
\end{eqnarray}
were
\begin{eqnarray}
\hat{S}_1\equiv S=\frac{\langle 1n\rangle}{\langle ns\rangle\langle s1\rangle },
~\hat{S}_2=
\frac{\tilde{\lambda}^{\dot{\alpha}}_s}{\langle s1\rangle}
\frac{\partial}{\partial\tilde{\lambda}^{\dot{\alpha}}_1}+
\frac{\tilde{\lambda}^{\dot{\alpha}}_s}{\langle sn \rangle}
\frac{\partial}{\partial\tilde{\lambda}^{\dot{\alpha}}_n}+
\frac{\eta_{A,s}}{\langle s1 \rangle}\frac{\partial}{\partial \eta_{A,1}}+
\frac{\eta_{A,s}}{\langle sn \rangle}\frac{\partial}{\partial \eta_{A,n}},
\end{eqnarray}
and $A$ is $SU(4)_R$ index. "Soft leg" $s$ may be in any position between legs $i$ and $i+1$ and we have chosen $i=n$ only for convenience. At the same time, while the behavior of form factor when one of the momenta associated with external particles
become soft is essentially identical to the amplitude case, its behavior in the limit
when the momentum of the operator $q$ becomes soft ($q$ and its Grassmann counterpart $\gamma$ $(q,\gamma) \mapsto 0$) is different. In fact, the following relation holds\footnote{See appendix \ref{aA} for notation.} (see \cite{BKV_SuperForm}):
\begin{equation}\label{cojectureAmpl-FF}
Z_{n}^{(k)}(\{\lambda_i,\tilde{\lambda}_i,\eta_i\};0,0)\sim g\frac{\partial
A_{n}^{(k)}(\{\lambda_i,\tilde{\lambda}_i,\eta_i\})}{\partial g},
\end{equation}
where $g$ is the coupling constant. It is interesting to note, that this relation must also hold at loop level.

The simple relation between $\mbox{MHV}$ form factors and amplitudes (\ref{MHV_formfactors_amplitudes}) suggests, that  on-shell diagrams for form factors will be identical to the on-shell diagrams for amplitudes with one of the external $\mbox{MHV}_3$ vertexes replaced with
\begin{eqnarray}
Z_2^{(2)}&=&\int  \frac{d\alpha_{1}}{\alpha_{1}}\frac{d\alpha_{2}}{\alpha_{2}}~Reg.(1,2|q)~
\delta^{2}\left(\tilde{\lambda}_1+\alpha_1\tilde{\lambda}_3\right)
\delta^{2}\left(\tilde{\lambda}_2+\alpha_2\tilde{\lambda}_3\right)
\times\nonumber\\
&\times&\delta^{2}\left(\lambda_3+\alpha_1\lambda_1+\alpha_2\lambda_2\right)
\hat{\delta}^{4}\left(\eta_1+\alpha_1\eta_3\right)
\hat{\delta}^{4}\left(\eta_2+\alpha_2\eta_3\right) ,
\end{eqnarray}
where we introduced the following notation for the inverse soft factor $S^{-1}$:
\begin{eqnarray}
Reg(i,i+1|q)\equiv S^{-1}(i,q,i+1)=\frac{\langle iq \rangle\langle q i+1\rangle}{\langle ii+1\rangle}.
\end{eqnarray}
Fig.\ref{5pointOnSellDiagramMHVFormF} shows the corresponding on-shell diagram in the case of $\mbox{NMHV}_4$ form factor (regulated vertex was denoted by red circle). The on-shell forms corresponding to such on-shell diagrams are then given by:
\begin{eqnarray}\label{regOnSellDiagram}
\Omega_n^{(k)}&=&\int \prod_{i=1}^{n_w}\frac{d\alpha_{1i}}{\alpha_{1i}}\frac{d\alpha_{2i}}{\alpha_{2i}}
 \prod_{j=1}^{n_b}\frac{d\beta_{1i}}{\beta_{1i}}\frac{d\beta_{2i}}{\beta_{2i}}
\prod_{m=1}^{n_I}\frac{1}{U(1)_m}Reg(l_1[\vec{\alpha}],l_2[\vec{\alpha}]|q)
 \times\nonumber\\
 &\times&\delta^{4|4}(1,\ldots,i,q,i+1,\ldots,n),
\end{eqnarray}
where
\begin{eqnarray}
&&\delta^{4|4}(1,\ldots,i,q,i+1,\ldots,n)=\nonumber\\ &=&\prod_{a=1}^k
 \delta^{2}\left(\sum_{l=1}^{n+1} C_{al}[\vec{\alpha}]\tilde{\lambda}_l\right)
 \delta^{4}\left(\sum_{l=1}^{n+1} C_{al}[\vec{\alpha}]\eta_l\right)\prod_{b=k+1}^n
 \delta^{2}\left(\sum_{l=1}^{n+1} \tilde{C}_{al}[\vec{\alpha}]\lambda_l\right),
\end{eqnarray}
and $\lambda$'s and $\eta$'s are taken from the ordered set $(1,\ldots,i,q,i+1,\ldots,n)$. The Weyl spinors $\lambda_{l_1}$ and $\lambda_{l_2}$ in $Reg$ function could be written as:
\begin{eqnarray}
\lambda_{l_i}=\sum_j\lambda_j a_j^i[\vec{\alpha}],
\end{eqnarray}
where $a_j^i[\vec{\alpha}]$ are dimensionless functions of coordinates on Grassmannian. The explicit form of such functions will in general depend on the on-shell diagram under consideration. In the following we will refer to the on-shell diagrams with $Reg.$ function included in one of its external vertexes as \emph{regulated on-shell diagrams}.

\begin{figure}[t]
 \begin{center}
  \epsfxsize=4cm
 \epsffile{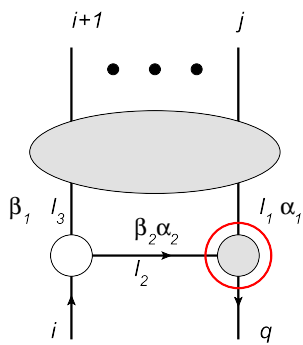}
 \end{center}\vspace{-0.2cm}
 \caption{BCFW bridge configuration with regulated vertex. }\label{BCFWBridgeOnShellRegDiagram}
 \end{figure}

To make expressions like (\ref{regOnSellDiagram}) useful from  the computational point
of view one must provide an algorithm for constructing  explicit form of $a_j^i[\vec{\alpha}]$ functions for a given on-shell diagram. For one class
of on-shell diagrams the form of $a_j^i[\vec{\alpha}]$ is particularly simple. These are
on-shell diagrams where regulated $\mbox{MHV}_3$
vertex with external leg $q$ is connected to the $\overline{\mbox{MHV}}_3$
vertex with external leg $i$ via so called BCFW bridge
(see Fig.\ref{BCFWBridgeOnShellRegDiagram} as an example with $i=3$). In this particular case one can choose individual edge variables such that $\mbox{MHV}_3$
and $\overline{\mbox{MHV}}_3$ vertexes become proportional to
\begin{eqnarray}
&&\mbox{MHV}_3:~
\delta^2(\tilde{\lambda}_{l_1}+\alpha_1\tilde{\lambda}_{q})
\delta^2(\tilde{\lambda}_{l_2}+\alpha_2\tilde{\lambda}_{q})
\delta^2(\lambda_q+\alpha_1\lambda_{l_1}+\alpha_2\lambda_{l_2}),\nonumber\\
&&\overline{\mbox{MHV}}_3:~\delta^2(\lambda_{l_3}+\beta_1\lambda_{i})
\delta^2(\lambda_{l_2}+\beta_2\lambda_{i})
\delta^2(\tilde{\lambda}_i+\beta_1\tilde{\lambda}_{l_3}+\beta_2\tilde{\lambda}_{l_2}) .
\end{eqnarray}
Solving the constraints  given by $\mbox{MHV}_3$
and $\overline{\mbox{MHV}}_3$ vertexes $\delta$-functions we get
\begin{eqnarray}
&&\mbox{MHV}_3: \langle l_1q\rangle=\alpha_2 \langle l_1l_2\rangle,\nonumber\\
&&\overline{\mbox{MHV}}_3: \lambda_{l_2}=\lambda_i\beta_2 ,
\end{eqnarray}
so that the $Reg.$ function is written as
\begin{eqnarray}\label{RegForBCFWBridge}
Reg(l_1[\vec{\alpha}],l_2[\vec{\alpha}]|q)
=\frac{\langle ql_1[\vec{\alpha}]\rangle\langle l_2[\vec{\alpha}]q \rangle}
{\langle l_1[\vec{\alpha}]l_2[\vec{\alpha}]\rangle}=\langle iq\rangle \alpha_2\beta_2.
\end{eqnarray}
Note, that the factors $\alpha_2\beta_2$ now cancel
with the similar factors from integration measure
$$
\prod_{i=1}^{n_w}\frac{d\alpha_{1i}}{\alpha_{1i}}\frac{d\alpha_{2i}}{\alpha_{2i}}
 \prod_{j=1}^{n_b}\frac{d\beta_{1i}}{\beta_{1i}}\frac{d\beta_{2i}}{\beta_{2i}}
\prod_{m=1}^{n_I}\frac{1}{U(1)_m},
$$
and will remove singular behavior with respect to $d\alpha_2d\beta_2$ integration. This further supports our initial idea that additional inverse soft factor associated with regulated $\mbox{MHV}_3$ vertex will regulate soft behavior of corresponding on-shell diagram with respect to some external momenta (with respect to the "$q$ leg" in our case). In the case of more general configurations of $\mbox{MHV}_3$ and $\overline{\mbox{MHV}}_3$ vertexes the explicit form of $Reg$ function will in principle be different and will depend on the choice of coordinates on the Grassmannian.

In the previous section we noted that in the case of amplitudes the on-shell diagram corresponding to the top cell on-shell form $\Omega^{top}$ is of particular interest. We have also mentioned that there are different possible choices of coordinates on Grassmannian and one of them is given by the elements of $C_{ai}$ matrix itself. Now also note, that in such coordinates $Reg$ function (\ref{RegForBCFWBridge}) for top cell diagram can be written as (at least for the simplest cases of on-shell diagrams
relevant for $\mbox{MHV}_n$ and $\mbox{NMHV}_4$ form factors)
\begin{eqnarray}
Reg(l_1,l_2|q)=\langle iq\rangle \frac{M_a}{M_b}, ~M_{a,b}~
\mbox{are some (in general non-consecutive) minors of matrix}~C_{ai}.\nonumber\\
\end{eqnarray}

The main goal of the present consideration of on-shell diagrams for form factors is to find an analog of $\Omega^{top}$ for $\mbox{N}^{k-2}\mbox{MHV}_n$ form factors. It is reasonable to suggest that an analog of $\Omega^{top}$ for form factors could be found
as a linear combination of regulated $\Omega^{top}$ on-shell forms for the amplitudes,
where $Reg$ functions are chosen in the form of ansatz
\begin{eqnarray}
Reg=\sum_i\langle iq\rangle \frac{M_a^{(i)}}{M_b^{(i)}}.
\end{eqnarray}
The explicit form of $M_a^{(i)}/M_b^{(i)}$ could be further fixed by comparison with some known explicit results from BCFW recursion. Indeed, it is easy to see, that in the case when Grassmannian integral is fully localized on $\delta$-functions, i.e. in the case of $\mbox{N}^{k-2}\mbox{MHV}_{k+1}$  (green arrow in Fig.\ref{ComputedFormFactorsPlot}) and $\mbox{MHV}_{n}$ form factors (red arrow in Fig.\ref{ComputedFormFactorsPlot}) the latter could be written  as linear combination of $\mbox{N}^{k-2}\mbox{MHV}_{k+2}$ and $\mbox{MHV}_{n+1}$ amplitudes:
\begin{figure}[t]
 \begin{center}
  \epsfxsize=7cm
 \epsffile{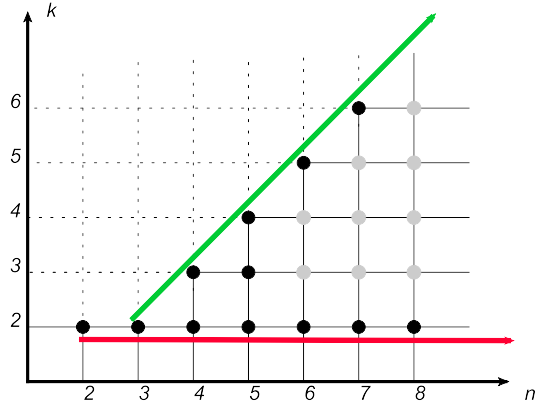}
 \end{center}\vspace{-0.2cm}
 \caption{Schematic representation of none zero form factors $Z^{(k)}_n$ with $q^2=0$. The black dots corresponds to the form factor reproduced in this paper.}
 \label{ComputedFormFactorsPlot}
 \end{figure}
\begin{eqnarray}\label{FFactorsNkMHVLinearCombAmpl}
Z^{(k)}_{k+1}&=&
\sum_{j=4}^{k+1}\frac{\langle q|p_1+p_j+\ldots+p_{k+1}|2]}
{[q2]}A^{(k)}_{k+2}(1,\ldots,j-1,q,j,\ldots,k+1)+\nonumber\\
&+&\frac{\langle q|p_1|2]}{[q2]}A^{(k)}_{k+2}(1,\ldots,k+1,q),~k\geq3.
\end{eqnarray}
and
\begin{eqnarray}
Z^{(2)}_n=\langle q1\rangle\frac{\langle qn\rangle}{\langle 1n \rangle} A^{(2)}_{n+1}(1,\ldots,n,q).
\end{eqnarray}
This representation could be obtained from BCFW recursion for $[1,2\rangle$ shift.
Analyzing coefficients in front of $A^{(k)}_{k+2}$  amplitudes as well as individual contributions to BCFW recursion in $\mbox{NMHV}$ sector we can fix the form
of $M_a^{(i)}/M_b^{(i)}$ minor ratios as well as  explicit form for the sum of regulated
on-shell forms $\Omega^{top}$ which should reproduce $\mbox{N}^{k-2}\mbox{MHV}_n$ form factors after integration over appropriate contours.

\begin{figure}[t]
 \begin{center}
  \epsfxsize=12cm
 \epsffile{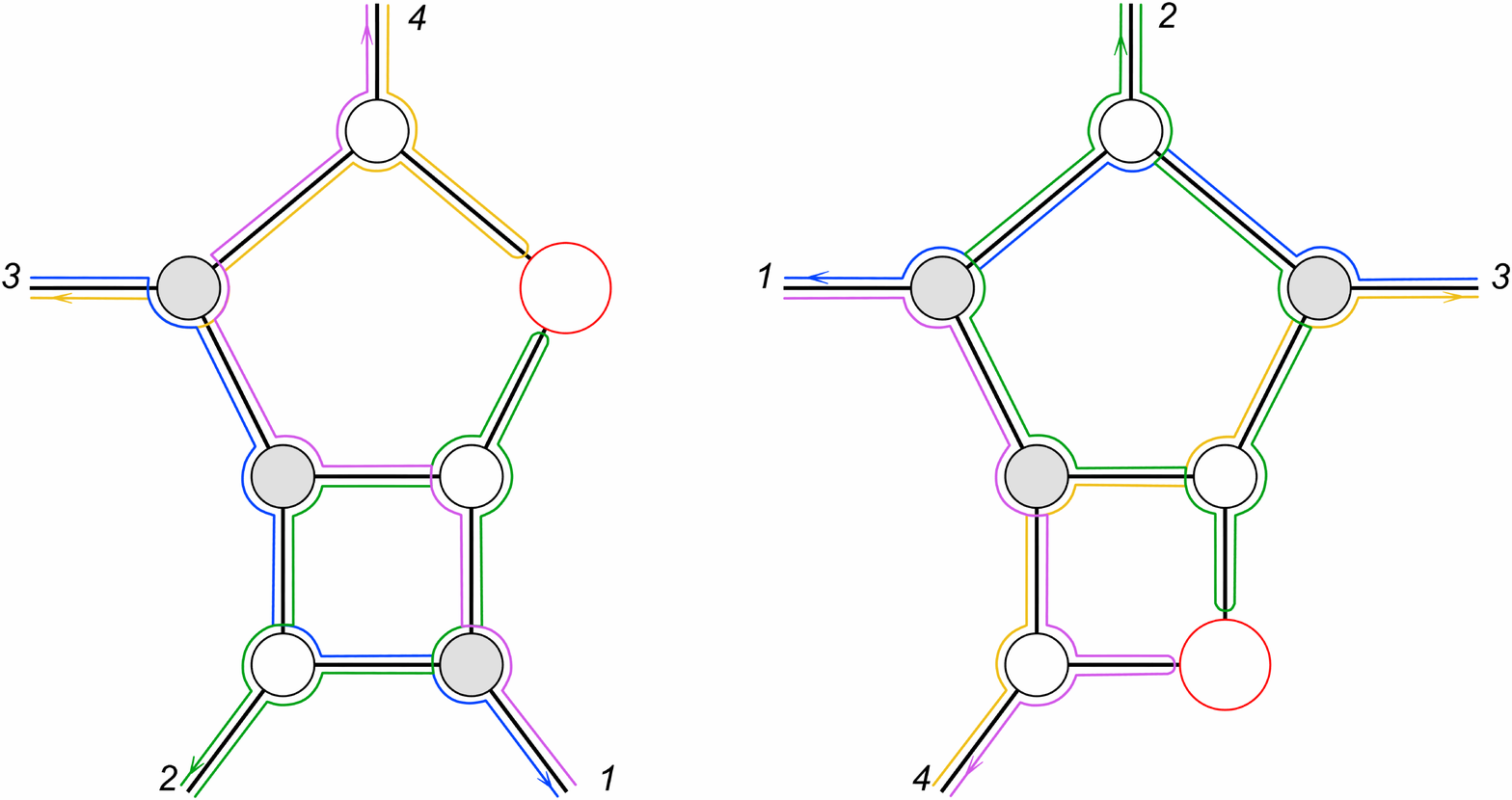}
 \end{center}
 \caption{Example of two on-shell diagrams with identical permutations but with different $Reg$
 functions. The regulated vertex corresponding to form factor is not shown.}\label{IndenticalPermRegOnSellDiagram}
 \end{figure}

The explicit results for  Grassmannian integral representation for form factors of operators from stress-tensor operator supermultiplet at $q^2=0$ will be given in next section, while at the end of this section we want to make some speculations about the role of permutations for regulated on-shell diagrams we introduced.
\begin{figure}[t]
 \begin{center}
  \epsfxsize=3cm
 \epsffile{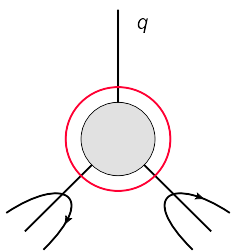}
 \end{center}\vspace{-0.2cm}
 \caption{Elementary permutation ("paths") associated with regulated $\mbox{MHV}_3$
 vertex in the on-shell diagrams. }\label{RegMHVVertex}
 \end{figure}
The permutation associated with a given on-shell diagram can be constructed by starting from external leg $i$ and moving along the
"left-right path" until finishing at another external leg $j$.
The natural prescription when there is regulated vertex in the on-shell diagram may
be the following: one should "turn back" at regulated vertex (see Fig. \ref{RegMHVVertex}). This way the regulated on-shell diagrams which differ from one another by the explicit form of $Reg$ function will correspond to the same permutation.
Then it is natural to conjecture that one must sum over such
sets of on-shell diagrams. This may explain why one have to consider linear combination
of top-cell like objects in the case of form factor in contrast to the amplitude case. See Fig. \ref{IndenticalPermRegOnSellDiagram} for example of on-shell diagrams relevant
to the $\mbox{NMHV}_4$ case.

\section{Conjecture for Grassmannian representation for form factors with $q^2=0$}\label{p4}
Now we are ready to present a conjecture for the analog of top-cell Grassmannian integral for form factors of operators from stress tensor operator supermultiplet at $q^2=0$. We claim that by appropriate choice of integration contour $\Gamma$ the
on-shell form $\Omega^{(k)}_n[\Gamma]$:
\begin{eqnarray}\label{GrassmannIntegralFormFactors}
\Omega^{(k)}_n[\Gamma]&=&\sum_{j=4}^{k+1}\int_{\Gamma} \frac{d^{n+1\times k}C_{al}}{Vol[GL(k)]}
~\frac{Reg^{R,(k)}_j}{M_1...M_{n+1}}~\delta^{4|4}(1,\ldots,j-1,q,j,\ldots,n)+\nonumber\\
&+&\int_{\Gamma} \frac{d^{n+1\times k}C_{al}}{Vol[GL(k)]}
~\frac{Reg^{L,(k)}_n}{M_1...M_{n+1}}~\delta^{4|4}(1,\ldots,n,q),
\end{eqnarray}
will reproduce all tree level
$\mbox{N}^{k-2}\mbox{MHV}$ form factors of operators from stress tensor supermultiplet
with $q^2=0$. Here, functions\footnote{Here superscripts $L$ and $R$ denote terms which give rise to BCFW contributions with the form factor standing either to the left or to the right of amplitude. See the discussion at the end of this section.} $Reg^{R,(k)}_j$ and $Reg^{L,(k)}_n$ regulate soft behavior of $\Omega^{(k)}_n[\Gamma]$ with respect to $\{\lambda_q,\tilde{\lambda}_q\}$ momentum and are given by
\begin{eqnarray}\label{RegFunctionsInGrassmannIntegral}
Reg^{R,(k)}_j&=&\langle q1\rangle\frac{(k+1k+2~3\ldots k)}{(13\ldots j-1~ j+1\ldots n+1)}
+\sum_{i=j}^{k+1}\langle qi\rangle\frac{(13\ldots i~i+2\ldots k+2)}{(13\ldots j-1~ j+1\ldots n+1)}
\nonumber\\
Reg^{L,(k)}_n&=&\langle q1\rangle\frac{(nn+1~3\ldots k)}{(1n~3\ldots k)},
\end{eqnarray}
for $k\geq3$ and by
\begin{eqnarray}
Reg^{R,(2)}_j=0,~Reg^{L,(2)}_n=\langle q1\rangle\frac{(nn+1)}{(1n)},
\end{eqnarray}
for $k=2$.
For example, the expressions for $\mbox{NMHV}_{4,5}$ form factors can be obtained using
$\Omega^{(3)}_4$ and $\Omega^{(3)}_5$ on-shell forms (for saving space we will use shorthand notation $\int \frac{d^{n+1\times k}C_{al}}{Vol[GL(k)]} \equiv \int$):
\begin{eqnarray}
\Omega^{(3)}_4=\int
\left(\langle1q\rangle \frac{(345)}{(135)}+
\langle4q\rangle \frac{(134)}{(135)} \right)\frac{\delta^{4|4}(1,2,3,q,4)}{M_1\ldots M_5}+
\int \left(\langle1q\rangle \frac{(345)}{(134)}\right) \frac{\delta^{4|4}(1,2,3,4,q)}{M_1\ldots M_5},\nonumber\\
\end{eqnarray}
\begin{eqnarray}\label{NMHV5FormFactorGrassmannIntegral}
\Omega^{(3)}_5=\int \left(\langle1q\rangle \frac{(345)}{(135)}+\langle4q\rangle
\frac{(134)}{(135)} \right)\frac{\delta^{4|4}(1,2,3,q,4,5)}{M_1\ldots M_6}
+\int \left(\langle1q\rangle \frac{(356)}{(135)}\right)
\frac{\delta^{4|4}(1,2,3,4,5,q)}{M_1\ldots M_6}.\nonumber\\
\end{eqnarray}
In the case of $\mbox{N}^{2}\mbox{MHV}_5$ form factor the corresponding expression could be obtained using $\Omega^{(4)}_5$ on-shell form:
\begin{eqnarray}
\Omega^{(4)}_5&=&\int \left(\langle1q\rangle \frac{(3456)}{(1356)}
+\langle4q\rangle \frac{(1346)}{(1356)}+\langle5q\rangle \frac{(1345)}{(1356)} \right)
\frac{\delta^{4|4}(1,2,3,q,4,5)}{M_1\ldots M_6}+\nonumber\\
&+&\int \left(\langle1q\rangle \frac{(3456)}{(1346)}+
\langle5q\rangle \frac{(1345)}{(1346)}\right)\frac{\delta^{4|4}(1,2,3,4,q,5)}{M_1\ldots M_6}+
\nonumber\\&+&
\int \left(\langle1q\rangle \frac{(3456)}{(1345)}\right) \frac{\delta^{4|4}(1,2,3,4,5,q)}{M_1\ldots M_6}.
\end{eqnarray}
Note that in all expressions above the integrations are made with respect to
$C_{ai}$ matrix elements parameterizing the points of corresponding Grassmannians. All the above expressions could be combined under one integral sign and were split into parts only for convenience. In the next sections we will present the checks of our conjecture on some particular examples as well as investigate different choices for  integration contours.

Before proceeding to the next section let us stop for the moment and discuss additional heuristic arguments in favor of our conjecture for the analog of top cell object for form factors. First, in the case of the amplitudes with $n=k+2$ there is only one contribution from BCFW recursion (at fixed $k$) coinciding with top cell on-shell diagram, so that the corresponding integration over
Grassmannian  is trivial and is fully localized on $\delta$-functions.
In the case of form factors with $q^2=0$ the analog of $n=k+2$ series for amplitudes
is given by $n=k-1$ series. However, there are now $k-1$ contributions from BCFW recursion (at fixed $k$). Each contribution is proportional to the regulated amplitude
like top cell on-shell diagram with regulated vertex with momentum $q$ being inserted between vertexes with momenta $i$ and $i+1$. The explicit positions of insertions in (\ref{GrassmannIntegralFormFactors}) may be related to permutations associated with regulated on-shell diagrams. We want to stress, that (\ref{GrassmannIntegralFormFactors}) reproduces $n=k+1$ series of form factors by construction. Next we assume that for fixed $k$ and $n>k+1$ (when Grassmannian integral is no longer localized on $\delta$-functions) both the structure of $Reg$ functions and their insertion positions will be essentially the same. In the next section we use the nontrivial example of $\mbox{NMHV}_5$ form factor to verify this claim.

Finally, the BCFW terms could be split into two groups with respect to whether the form factor stands to the left or to the right of the amplitude in the corresponding BCFW diagram. This explains $R$ and $L$ superscript notation in $Reg$ functions. Then,
the residues of corresponding Grassmannian integrals should reproduce "left" and "right"
BCFW terms.

\section{$\mbox{MHV}_n$,$~\mbox{N}^{k-2}\mbox{MHV}_{k+1}$, $\mbox{NMHV}_{5}$ form factors from Grassmannian integral and soft limit consistency check}\label{p5}

The $\mbox{MHV}_n$ form factors (red arrow in Fig. \ref{ComputedFormFactorsPlot}):
 \begin{eqnarray}
Z^{(2)}_n=\langle q1\rangle\frac{\langle qn\rangle}{\langle 1n \rangle} A^{(2)}_{n+1}(1,\ldots,n,q)=\frac{\delta^8(q_{1\ldots n}+\gamma)}{\langle12\rangle \ldots \langle n1\rangle},
\end{eqnarray}
are reproduced from (\ref{GrassmannIntegralFormFactors}) trivially. For a series of form factors with fixed $k$ and $n=k+1$ we should verify that integration over Grassmannian in (\ref{GrassmannIntegralFormFactors}) reproduces explicit results (\ref{FFactorsNkMHVLinearCombAmpl}) following from  BCFW recursion. We have explicitly checked that in the case of  $Z^{(3)}_{4}$, $Z^{(4)}_{5}$ and $Z^{(5)}_{6}$ form factors both BCFW recursion with $[1,2\rangle$ shift and Grassmannian representation (\ref{GrassmannIntegralFormFactors}) give
\begin{figure}[t]
 \begin{center}
  \epsfxsize=14cm
 \epsffile{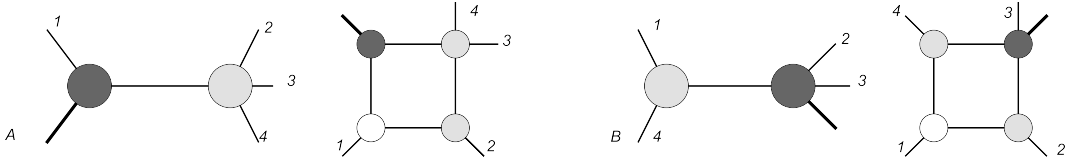}
 \end{center}\vspace{-0.2cm}
 \caption{BCFW diagrams for $[1,2\rangle$ shift and corresponding $R_{ijk}^{(1,2)}$ functions for $Z^{(3)}_4$. Dark grey vertex is $\mbox{MHV}_n$ form factor.}\label{BCFWdiagramsNMHV4}
 \end{figure}
 \begin{figure}[t]
 \begin{center}
  \epsfxsize=15cm
 \epsffile{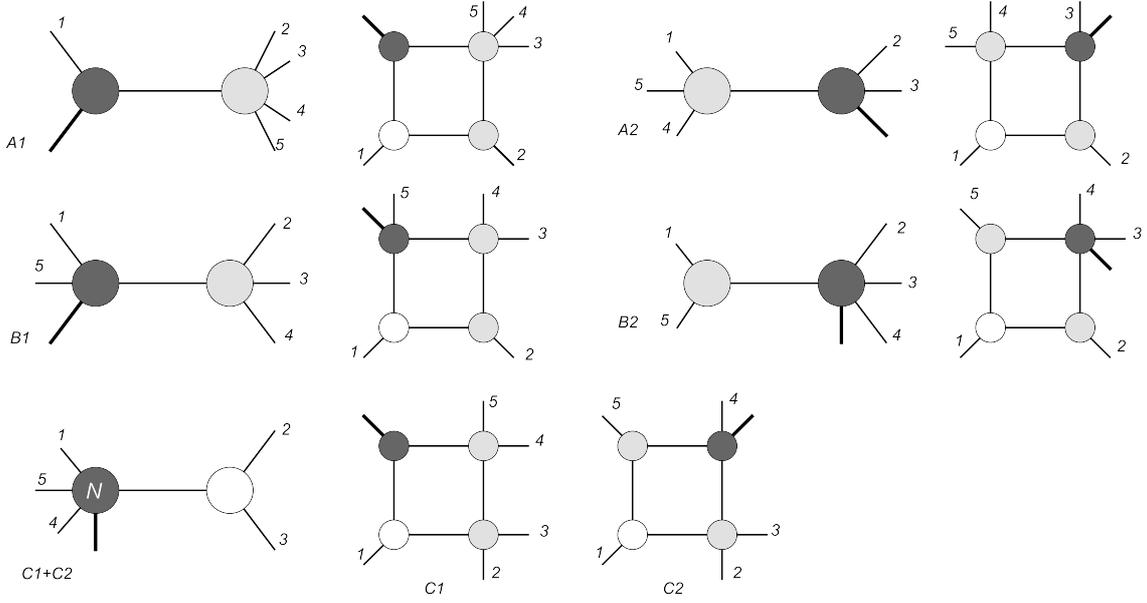}
 \end{center}\vspace{-0.2cm}
 \caption{BCFW diagrams for $[1,2\rangle$ shift and corresponding $R_{ijk}^{(1,2)}$ functions for $Z^{(3)}_5$. Dark grey vertexes are $\mbox{MHV}_n$ and $\mbox{NMHV}_4$ form factors.}\label{BCFWdiagramsNMHV5}
 \end{figure}
\begin{eqnarray}\label{ZNMHV4}
Z^{(3)}_{4}&=&\delta^8(q_{1\ldots4}+\gamma)\frac{\hat{\delta}^4(123)}{\langle4q\rangle^4}
\left(\frac{\langle1q\rangle[12][q3][14]+
\langle3q\rangle[23][34][q1]}{[1q][2q][3q][4q]\mathcal{P}^*(1234)}\right),
\end{eqnarray}
and
\begin{eqnarray}\label{ZNNMHV5}
Z^{(4)}_{5}&=&\delta^8(q_{1\ldots5 }+\gamma)\Big{(}
\frac{\hat{\delta}^4(12q)\hat{\delta}^4(345)}{(p^2_{345})^4}
\frac{\langle q|p_1|2]}{[q2]\mathcal{P}^*(12345q)}
+\frac{\hat{\delta}^4(125)\hat{\delta}^4(34q)}{(p^2_{125})^4}
\frac{\langle q|p_3|5]}{[q5]\mathcal{P}^*(12q345)}\nonumber\\
&+&\frac{\hat{\delta}^4(125)\hat{\delta}^4(34q)}{(p^2_{125})^4}
\frac{\langle q|p_4|2]}{[q2]\mathcal{P}^*(1234q5)}\Big{)}.
\end{eqnarray}
Here, to have a compact representation we introduced some new notation which is explained in appendix \ref{aA}. The details of Grassmannian integral evaluation as well as explicit results for $Z^{(5)}_{6}$ form factor could be found in appendix \ref{aB}.  In the case of $Z^{(3)}_{4}$, $Z^{(4)}_{5}$ and $Z^{(5)}_{6}$ form factors we have also
verified cyclical symmetry of the result with respect to permutation $\mathbb{P}$ of
momenta of external particles (the permutation does not act on the momentum $q$ of operator)

Next, we verified that (\ref{GrassmannIntegralFormFactors}) reproduces BCFW result for
$Z^{(3)}_{5}$ form factor, which is none trivial check as the result for this form factor was not used when deriving (\ref{GrassmannIntegralFormFactors}). The BCFW result for $Z^{(3)}_{5}$ form factor contains 6 terms, which could be extracted either
from the general solution of BCFW recursion in $\mbox{NMHV}$ sector or from direct consideration of $[1,2\rangle$ BCFW shift for this particular form factor
(see Fig. \ref{BCFWdiagramsNMHV5} and (\ref{SolutionOfBCFWNMHVsector})):
\begin{eqnarray}\label{SolutionOfBCFWNMHVsector}
    Z_{n}^{(3)}=Z_{n}^{(2)}\left(\sum_{i=2}^{n-2}\sum_{j=i+1}^{n-1}R^{(1)}_{1ji}+
    \sum_{i=2}^{n-2}\sum_{j=i+2}^{n}R^{(2)}_{1ji}\right),
\end{eqnarray}
where the definition of $R_{rst}^{(1,2)}$ functions could be found in appendix \ref{aB}. We will label mentioned six terms as $A1=Z_{n}^{(2)}R^{(2)}_{152}$,
$B1=Z_{n}^{(2)}R^{(2)}_{142}$, $C1=Z_{n}^{(2)}R^{(2)}_{153}$
and $A2=Z_{n}^{(2)}R^{(2)}_{132}$, $=Z_{n}^{(2)}R^{(2)}_{142}$,
$C2=Z_{n}^{(2)}R^{(2)}_{153}$ (see Fig. \ref{BCFWdiagramsNMHV5}). The explicit expressions for these terms are given in appendix \ref{aB}. So, we have
\begin{eqnarray}
Z^{(3)~[1,2\rangle}_{5}=A1+B1+A2+B2+C1+C2.
\end{eqnarray}
The integral over Grassmannian in this case is no longer localized on $\delta$-functions and can be reduced to one-dimensional integral over complex parameter $\tau$, which could be further evaluated by residues. It is convenient to label the residues of integral at poles $1/M_i$ and $1/(135)$ as $\{i\}$ and $\{*\}$ correspondingly. We also choose the contour of integration over $\tau$ $\Gamma_{135}$
to encircle poles $\{5\},\{3\}$ and $\{1\}$ similar to the amplitude case. This way we get
\begin{eqnarray}
\{5\}=B1+B2,~\{3\}=A1+A2,~\{1\}=C1+C2,
\end{eqnarray}
and
\begin{eqnarray}
\Omega^{(3)}_5[\Gamma_{135}]=Z^{(3)~[1,2\rangle}_{5}.
\end{eqnarray}
It is interesting to note that if we split the Grassmannian integral into "left" and "right" parts then
$A1,B1$ terms will be given by "left",  while $A2,B2$ terms by "right" part. That is
\begin{eqnarray}
A1+B1=\int_{\Gamma_{53}} Reg^{L,(3)}_5\frac{\delta^{4|4}(1,2,3,4,5,q)}{M_1\ldots M_6},
~A2+B2=\int_{\Gamma_{53}} Reg^{R,(3)}_4 \frac{\delta^{4|4}(1,2,3,q,4,5)}{M_1\ldots M_6}.
\nonumber\\
\end{eqnarray}
For the residues at $\{1\}$ pole on the other hand we get
\begin{eqnarray}
\tilde{C}1=\int_{\Gamma_{1}} Reg^{L,(3)}_5\frac{\delta^{4|4}(1,2,3,4,5,q)}{M_1\ldots M_6},
~\tilde{C}2=\int_{\Gamma_{1}} Reg^{R,(3)}_4 \frac{\delta^{4|4}(1,2,3,q,4,5)}{M_1\ldots M_6}.
\end{eqnarray}
where individual terms $\tilde{C}1$ and $\tilde{C}2$ are different from $C1,C2$, but fortunately their sums coincide $\tilde{C}1+\tilde{C}2=C1+C2$. From this particular example we see that analytical relations between individual BCFW contributions and individual residues of $\Omega^{(k)}_n$ become rather none trivial even in
$\mbox{NMHV}$ sector in contrast to the amplitude case.

Let us now perform another self consistency check of our conjecture. It was initially claimed that $R^{R,(k)}_j$ and $R^{L,(k)}_n$ functions should regulate soft behavior of form factors with respect to soft limit  $q \rightarrow 0$. The soft behavior of amplitudes within Grassmannian integral formulation was considered in details in \cite{SoftTheoremsGrassmannian}. Here we want to use the results of \cite{SoftTheoremsGrassmannian} to show that the relation
(\ref{cojectureAmpl-FF}) could be also reproduced by taking soft limit with respect to momentum  $q$ in (\ref{GrassmannIntegralFormFactors}). In other words  if our conjecture for $\Omega^{(k)}_n$ in the case of form factors is correct then the following relation must hold\footnote{Here $\sim$ means the presence of numerical coefficient $k-1$.}:
\begin{eqnarray}
	\Omega^{(k)}_n[\Gamma_n^{tree}]\Big{|}_{\lambda_q\mapsto \epsilon \lambda_q}\sim A_n^{(k)}+O(\epsilon),~\epsilon\rightarrow 0.
\end{eqnarray}
Here $\Gamma_n^{tree}$ is the contour corresponding to $\mbox{N}^{k-2}\mbox{MHV}_n$ amplitude. For this purpose lets consider first non-trivial case given by $\Omega^{(3)}_n$ on-shell form. It is convenient to split it into left $\Omega^{L,(3)}_n$ and
right $\Omega^{R,(3)}_n$ parts. Lets consider $\Omega^{L,(3)}_n$ part first. Using the notation from \cite{SoftTheoremsGrassmannian} we parametrize $C_{al}$ matrix as (the columns are numerated as $(1,2,3,\ldots,n-1,n,n+1)$)
\begin{eqnarray}
    C=\left( \begin{array}{ccccccc}
        0 & c_{n-22} & a &\ldots& 1 & 0 & c_{n-2n} \\
        0 & c_{n-12} & b &\ldots& 0 & 1 & c_{n-1n} \\
        1 & c_{12}   & c &\ldots& 0 & 0 & c_{1n}\end{array} \right).
\end{eqnarray}
In this parametrization the minors in $Reg_n^{L,(3)}$ function are given by
\begin{eqnarray}
    (3nn+1)=ac_{1n},~(13n)=a,~\mbox{so}~\frac{(3nn+1)}{(13n)}=c_{1n}.
\end{eqnarray}
and $\Omega^{L,(3)}_n$ in the vicinity of point $(nn+11)=0$ could be written as
\begin{eqnarray}
   \int_{\Gamma}
\frac{Reg^{L,(3)}_n\delta^{4|4}(1,\ldots,n,q)}{M_1...M_{n+1}}\Big{|}_{\lambda_q\mapsto \epsilon \lambda_q}&=&
 \int d^3c_{In}\delta^2(\epsilon\lambda_q-\lambda_Ic_{In})\frac{\epsilon Reg^{L,(3)}_n(n-1n1)'(n12)'}{(n-1nn+1)(nn+11)(n+112)}
\nonumber\\
&\times&
\int'_{\Gamma'} \frac{\delta^{4|4}(\hat{1},\ldots,n-1,\hat{n})}{M_1...M_{n}}\Big{|}_{\lambda_q\mapsto \epsilon \lambda_q},
\end{eqnarray}
where
\begin{eqnarray}
   \int\equiv \int \frac{d^{n+1\times 3}C_{al}}{Vol[GL(3)]},~
   \int'\equiv \int \frac{d^{n\times 3}C_{al}}{Vol[GL(3)]},~
   \int d^3c_{In}=\int dc_{1n}dc_{n-1n}dc_{n-2n},
\end{eqnarray}
and primes after some minors like $(n-1n1)'$ mean that they should be evaluated in  $Gr(3,n)$ Grassmannian compared to other minors evaluated in $Gr(3,n+1)$ Grassmannian,
$\Gamma$ contour contains the same poles as $\Gamma'$ plus additional pole $(nn+11)$. Extra hats, like
$\hat{1}$ and $\hat{n}$ mean that corresponding antiholomorphic spinors $\tilde{\lambda}_1$,$\tilde{\lambda}_n$ and $\eta_n$ get shifted as
\begin{eqnarray}\label{SoftShift}
   \hat{\tilde{\lambda}}_1&=&\tilde{\lambda}_1+c_{1n}\tilde{\lambda}_q,\nonumber\\
   \hat{\tilde{\lambda}}_n&=&\tilde{\lambda}_n+c_{n-1n}\tilde{\lambda}_q,\nonumber\\
   \hat{\eta}_n&=&\eta_n+c_{n-1n}\eta_q.
\end{eqnarray}
The sum $\lambda_Ic_{In}$ is given by
\begin{eqnarray}
   \lambda_Ic_{In}=\lambda_{n-1}c_{n-2n}+\lambda_{n}c_{n-1n}+\lambda_1c_{1n}.
\end{eqnarray}
The integral $\int d^3c_{In}$ is evaluated taking residue at pole $(nn+11)$, which
fixes the $c_{n-2n}$ and $c_{n-1n}$, $c_{1n}$ coefficients to be
\begin{eqnarray}\label{CcoefficientsInSoftLimit}
   c_{n-2n}=0,~c_{n-1n}=\frac{\langle 1q \rangle \epsilon}{\langle 1n \rangle},
   ~c_{1n}=\frac{\langle nq \rangle \epsilon}{\langle 1n \rangle}.
\end{eqnarray}
All other coefficients of $C_{la}$ matrix cancel out. Then the result of integration could be written as
\begin{eqnarray}
\int d^3c_{In}\delta^2(\epsilon\lambda_q-\lambda_Ic_{In})\frac{\epsilon Reg^{L,(3)}_n(n-1n1)'(n12)'}{(n-1nn+1)(nn+11)(n+112)}=
\frac{\langle 1n\rangle}{\epsilon^2\langle 1q \rangle \langle qn\rangle}
~\epsilon Reg^{L,(3)}\Big{|}_{(nn+11)},\nonumber\\
\end{eqnarray}
with $Reg^{L,(3)}_n$ evaluated at $(nn+11)$ given by
\begin{eqnarray}
\epsilon Reg^{L,(3)}\Big{|}_{(nn+11)}=
\frac{\epsilon^2\langle 1q \rangle \langle qn\rangle}{\langle 1n\rangle},
\end{eqnarray}
which is exactly inverse soft factor $S^{-1}(1,q,n)$ as we expected. So taking $\epsilon \rightarrow 0$
limit and taking into account (\ref{SoftShift}) and (\ref{CcoefficientsInSoftLimit}) we can write
\begin{eqnarray}
\Omega^{L,(3)}_n\Big{|}_{\lambda_q\mapsto \epsilon \lambda_q}=\int_{\Gamma}
\frac{Reg^{L,(3)}_n\delta^{4|4}(1,\ldots,n,q)}{M_1...M_{n+1}}\Big{|}_{\lambda_q\mapsto \epsilon \lambda_q}=A_n^{(3)}(1,\ldots,n)+O(\epsilon),
\end{eqnarray}
for $\Gamma =\Gamma_n^{tree}$.

Now lets turn to $\Omega^{R,(3)}_n$ contribution . Rearranging external kinematical data  such that $\delta^{4|4}(1,2,3,q,4,\ldots,n)=\delta^{4|4}(4,5,6,\ldots,n,1,2,3,q)$ and using the results obtained above with simple column relabeling in minors
\begin{eqnarray}
\begin{matrix}
1&2&3&\ldots&n-1&n&n+1\\
\downarrow&\downarrow&\downarrow&\ldots&\downarrow&\downarrow&\downarrow\\
5&6&7&\ldots&2&3&4
\end{matrix}
\end{eqnarray}
to evaluate their ratios
\begin{eqnarray}
\frac{(345)}{(135)}\Big{|}_{(345)}=0,~\frac{(134)}{(135)}=c_{1n},
\end{eqnarray}
together with the value of  $Reg_4^{R,(3)}$ function at $(345)$ residue
\begin{eqnarray}
	\epsilon Reg_4^{R,(3)}\Big{|}_{(345)}=
	\frac{\epsilon^2\langle 3q \rangle \langle q4\rangle}{\langle 34\rangle}.
\end{eqnarray}
we get
\begin{eqnarray}
\Omega^{R,(3)}_n\Big{|}_{\lambda_q\mapsto \epsilon \lambda_q}=\int_{\Gamma}
\frac{Reg^{R,(3)}_4\delta^{4|4}(1,2,3,q,5,\ldots,n)}{M_1...M_{n+1}}\Big{|}_{\lambda_q\mapsto \epsilon \lambda_q}=A_n^{(3)}(1,\ldots,n)+O(\epsilon),
\end{eqnarray}
with $\Gamma =\Gamma_n^{tree}$. Combining both contributions together we get
\begin{eqnarray}
\Omega^{(3)}_n\Big{|}_{\lambda_q\mapsto \epsilon \lambda_q}=2A_n^{(3)}(1,\ldots,n)+O(\epsilon).
\end{eqnarray}
Similar consideration for the case of $\Omega^{(k)}_n,~k>3$ is more complicated (but still possible using the results of \cite{SoftTheoremsGrassmannian}).
Most of the ratios of  minors in $Reg_j^{R,(k)}$ function should evaluate to $0$
due to specific gauge choice made when evaluating residues at corresponding poles. The rest of minors should evaluate to the $S^{-1}(j,q,j+1)$. The same should be true for $Reg_n^{L,(k)}$ function.

In the end of this section we would like to comment on the freedom in the choice for explicit $Reg^{L,(k)}_n$ and $Reg^{R,(k)}_j$ functions expressions. Most likely the choice made in (\ref{RegFunctionsInGrassmannIntegral}) is not unique. Indeed, we have the following curious identity for the minors ratios in the case of $\mbox{NMHV}_4$ and $\mbox{NMHV}_5$ form factors:
\begin{eqnarray}
	\langle 3q\rangle \frac{(145)}{(135)}=\langle 1q\rangle \frac{(345)}{(135)}+
	\langle 4q\rangle \frac{(134)}{(135)},
\end{eqnarray}
in the case of $\mbox{NMHV}_4$ form factor it is just the consequence of momentum conservation. Surprisingly the same relation holds also for the $\mbox{NMHV}_5$ form factor in a sense that
\begin{eqnarray}\label{RelationAmongMinors}
	\langle 3q\rangle \frac{(145)}{(135)}\Big{|}_{M_1,M_3,M_5}=\langle 1q\rangle \frac{(345)}{(135)}\Big{|}_{M_1,M_3,M_5}+
	\langle 4q\rangle \frac{(134)}{(135)}\Big{|}_{M_1,M_3,M_5},
\end{eqnarray}
where subscript $|_{M_1,M_3,M_5}$ means that minors should be evaluated at corresponding residues. This means that choosing
\begin{eqnarray}
	Reg^{R,(3)}_4=\langle 3q\rangle \frac{(145)}{(135)},
\end{eqnarray}
instead of (\ref{RegFunctionsInGrassmannIntegral}) will give us identical result
for $\mbox{NMHV}_4$ and $\mbox{NMHV}_5$ form factors.
One may wonder if relations like (\ref{RelationAmongMinors}) exist in the general $Gr(k,n)$
case\footnote{Note that these relations among minors are not, at least explicitly, Plucker relations since they involve
dimensionfull parameters such as $\langle iq\rangle$.}
and whether is it possible to simplify representation
(\ref{RegFunctionsInGrassmannIntegral}) for $Reg$ functions further.
We haven't found more simple expression that correctly reproduces
$\mbox{N}^{k-2}\mbox{MHV}_{k+1}$ and $\mbox{N}\mbox{MHV}_{n}$
form factors in a universal way, but of course that doesn't mean that such more simple
representation doesn't exists.

\section{Different contours in Grassmannian and $\mbox{NMHV}_{5}$ form factor}\label{p6}
\begin{figure}[t]
 \begin{center}
  \epsfxsize=5cm
 \epsffile{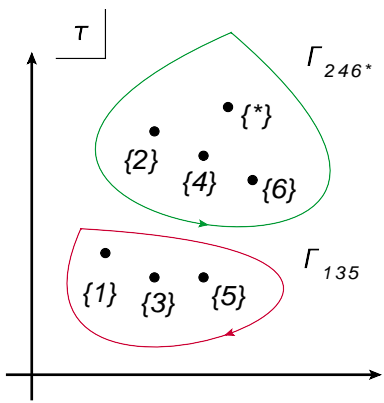}
 \end{center}\vspace{-0.2cm}
 \caption{Different integration contours in $\tau$ plain. }\label{DifferentIntegrationContours}
 \end{figure}
In this section we would like to discuss how the cancellation of spurious poles and the relations between different
BCFW representations for form factors follow from  our Grassmannian representation. To do that we will consider $\mbox{NMHV}_{5}$ form factor  discussed previously as an example.

Lets start with the relations between different BCFW representations for form factors. The general analytical structure of tree level form factors could be described as follows: the form factor is given by a sum of terms each having physical poles corresponding to different factorization channels. At the same time spurious poles if present should cancel in the sum of terms. In the case of $\mbox{NMHV}_{5}$ form factor the physical poles are either
of the form $\langle i i+1\rangle [i i+1]$ (so called collinear poles),
$\langle i q\rangle [i q]$ or of the
form $p_{ijk}^2=(p_i+p_j+p_k)^2$ (multiparticle poles). Here we will stop on the structure of the multiparticle poles.
For the $[1,2\rangle$ BCFW shift representation of $\mbox{NMHV}_{5}$ form factor they could be identified term by term, after some algebra, with the terms from the sum of residues $\{1\},\{3\},\{5\}$
in (\ref{NMHV5FormFactorGrassmannIntegral}) also having multiparticle poles (here we write $p_{ijk}^2$ with indexes matching those in corresponding Grassmann $\delta$ - functions
$\hat{\delta}^4(ijk)$):
\begin{eqnarray}
 \{1\}:~ p^2_{q45};~\{3\}:~ p^2_{23q},p_{12q};~\{5\}:~p^2_{234},p^2_{125};
\end{eqnarray}
that is $[1,2\rangle$ BCFW shift representation of $\mbox{NMHV}_{5}$
form factor contains the following set of multiparticle poles
\begin{eqnarray}
 P^{[1,2\rangle}=\{p^2_{45q},p^2_{23q},p_{12q},p^2_{234},p^2_{125}\}.
\end{eqnarray}
Using $[2,3\rangle$ BCFW shift one can obtain representation for $\mbox{NMHV}_{5}$ form factor with poles
\begin{eqnarray}
 P^{[2,3\rangle}=\{p^2_{23q},p^2_{34q},p_{345},p^2_{123},p^2_{51q}\}=\mathbb{P}P^{[1,2\rangle}.
\end{eqnarray}
It is easy to see that other BCFW representations will not contain new multiparticle poles as
$\mathbb{P}^2P^{[1,2\rangle}=P^{[1,2\rangle}$ and the set of poles
\begin{eqnarray}
 \{P^{[1,2\rangle},P^{[2,3\rangle}=\mathbb{P}P^{[1,2\rangle}\},
\end{eqnarray}
is closed under permutation $\mathbb{P}$. It is tempting to try to reproduce analytical expression for $[2,3\rangle$ BCFW shift representation of $\mbox{NMHV}_{5}$ form factor  as the sum
over residues given by contour $\Gamma_{246*}$\footnote{We want to emphasize that we are interested in explicit analytical relation between different BCFW representations. Otherwise, of course Cauchy theorem ensures that the sums of residues given by contours $\Gamma_{135}$ and  $\Gamma_{246*}$ are equal.}. Unfortunately such term by term identification is not possible without extra algebra involving rearrangements of spinor products (which is not surprising since term by term identification of $[1,2\rangle$ BCFW shift representation with the sum of residues corresponding to
contour $\Gamma_{135}$ already involves some algebra). On other hand the set of multiparticle poles in the sum of residues for the contour $\Gamma_{246*}$ is precisely given by $P^{[2,3\rangle}$. The collinear poles are identical in all BCFW representations/sums of residues. In any case we see that different choices of integration contours in our deformed Grassmannian integral representation allow us to obtain some non-trivial relations between rational functions similar to those in the amplitude case.

\begin{figure}[t]
 \begin{center}
  \epsfxsize=5cm
 \epsffile{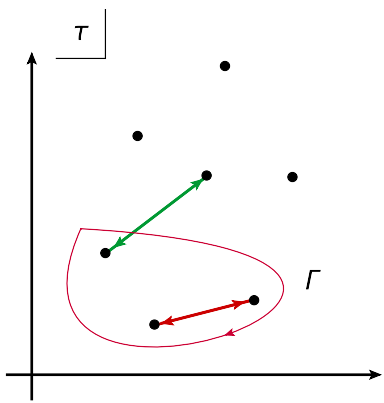}
 \end{center}\vspace{-0.2cm}
 \caption{Differences between physical and spurious poles. Red arrow corresponds to the "collision of spurious poles". Green arrow corresponds to the "collision of physical poles".}\label{DifferentIntegrationContoursPhysVsUnphys}
 \end{figure}
The careful reader may already noticed that the discussion of the relations between different BCFW representations is somewhat redundant (at least in the NMHV case), because momentum conservation  in this case allows one to rewrite the set of poles $P^{[1,2\rangle}$ in a
manifestly cyclically invariant form. That is the relation between different BCFW representations for form factors with $q^2=0$ may turn out to be trivial.

Let us now discuss the cancellation of spurious poles between individual BCFW terms contributing to $\mbox{NMHV}_{5}$ form factor. The situation here is identical to the case of $\mbox{NMHV}_6$ amplitude. The positions of $1/M_1,\ldots, 1/M_6$
and $1/(135)$ poles in complex $\tau$ plane depend on external kinematical data. The vanishing of some combinations of spinors like $p^2_{123} \rightarrow 0$ or $[3|4+5|q\rangle \rightarrow 0$ corresponds to the \emph{collisions} of two poles from the set  $1/M_1,\ldots, 1/M_6$, $1/(135)$. The difference between vanishing of $[3|4+5|q\rangle \rightarrow 0$ (which is a spurious pole of the individual BCFW term) and vanishing of $p^2_{123} \rightarrow 0$ (which is the physical pole of the form factor) is the following. In the case of  $[3|4+5|q\rangle \rightarrow 0$ the sum of residues for the Grassmannian integral with contour  $\Gamma_{135}$ (or $\Gamma_{246*}$) is always regular as the collision of poles occurs inside the integration contour and it is always possible to
choose opposite direction for it to avoid this possible singularity.
On other hand the situation with physical pole of the form factor (like $p^2_{123} \rightarrow 0$) is different and corresponds to the collision of  $\tau$ plane poles lying on the opposite sides of integration contour. In the case of $p^2_{123} \rightarrow 0$ we have the collision of $1/M_1$ and $1/M_4$ poles and this singularity can not be avoided (see Fig. \ref{DifferentIntegrationContoursPhysVsUnphys}). We expect that
similar situation will occur also in more complicated cases with $\mbox{N}^{k-2}\mbox{MHV}$ form factors in full analogy with the amplitude case.  This brings us to the following  questions: is it possible to interpret the residues of
(\ref{GrassmannIntegralFormFactors}) as a basis for the leading singularities of form factors and  whether there is a general prescription for the choice of integration contour in  more complicated cases of $\mbox{N}^{k-2}\mbox{MHV}$ form factors? As we have seen in the case of $\mbox{NMHV}_5$ form factor at least some of the residues are equal to the combination $Z_5^{(0)}R^{(1,2)}_{ijk}$.
The quadruple cuts of one-loop form factor will contain exactly this combination \cite{BORK_NMHV_FF}. However, mainly because there are no explicit answers available for the higher loop $\mbox{N}^{k-2}\mbox{MHV}$ form factors it is hard to speculate further. We are going to investigate this question in upcoming publications. One can also notice that the $\Gamma_{135}$ contour is in fact identical to the one in the case of $\mbox{NMHV}_6$ amplitude. We may conjecture that in the general case the
integration contour appropriate for the $\mbox{N}^{k-2}\mbox{MHV}_n$ form factors may be chosen similar to the case of $\mbox{N}^{k-2}\mbox{MHV}_{n+1}$ amplitude ($[1,2\rangle$ BCFW representation).

\section{Discussion and open questions}\label{p7}
Here we want to address several general questions regarding the  construction presented in this article and form factors of 1/2-BPS operators in $\mathcal{N}=4$ SYM in general.

First, it would be important to deeper understand the combinatorics behind introduced here regulated on-shell diagrams (the role of permutations, nonplanarity and so on). Among other things this may be important for the construction of the analog of BCFW recursion for the integrands of form factors at loop level.

Second, it would be interesting to investigate further soft limit properties of the form factors of operators from stress tensor operator supermultiplet with $q^2\neq0$. One should be able to recover (\ref{cojectureAmpl-FF}) via double soft limit with respect to the spinor variables parameterizing off-shell momentum $q$. The behavior of more general 1/2-BPS form factors is also likely to be regular with respect to $q \rightarrow 0$ limit. One can try to use the idea of regulated Grassmannian integral to describe form factors of these more general operators via the introduction of appropriate regulator functions similar to those in the case of $\mbox{MHV}_n$ 1/2-BPS form factors. However in the light of recent developments \cite{Wilhelm_Grassmannians_Integrability} - it is not clear whether this strategy is easier.

In \cite{Wilhelm_Grassmannians_Integrability} it was noted that at least in $\mbox{NMHV}$ sector one can separate the residues of Grassmannian integral (more accurately the ratio of Grassmannian integral and $\mbox{MHV}_n$ form factor) in two groups. Using momentum twistor representation one can show that one group contains residues proportional to
$$
A_i=c_i[1,n,i,\widehat{n+1},\widehat{n+2}],
$$
where $c_i$ is some rational function
of momentum twistor products $\langle abcd\rangle$, $\hat{\mathcal{Z}}_{n+1}$ and $\hat{\mathcal{Z}}_{n+2}$
twistors are introduced to close the period of the periodical contour in momentum twistor space, $i=2,...,n-1$. See \cite{Wilhelm_Grassmannians_Integrability} for details. The other group of residues are given by explicitly Yangian invariant functions
$$
B_{ijk}=[1,n, i,j,k],
$$
$i,j,k=2,...,n-1,\widehat{n+1},\widehat{n+2}$. So, in principle, one can always choose a contour of integration in such a way that to obtain Yangian invariant expression (the contour of integration which encircles only poles giving $B_{ijk}$ residues). However, such "Yangian invariant contour" will not lead to \emph{local} expressions and spurious poles will not cancel. In more complicated cases of ($\mbox{N}^{k-2}\mbox{MHV},~k>3$) form factors the situation is less clear. One may hope however, that the $q^2=0$ case is both simpler and "better" in this respect. We hope that in this case the integration contour in the Grassmannian integral may be chosen in a way, that both
Yangian invariance and locality will be preserved. Indeed, it is likely that regardless of the particular momentum twistor parametrization
the $\mbox{N}\mbox{MHV}_n$ form factors with $q^2=0$ are given by linear combination of $[a,b,c,d,e]$ Yangian invariants \cite{SoftTheoremsGrassmannian} (more accurately the ratio of $\mbox{N}\mbox{MHV}_n$ to $\mbox{MHV}_n$). We are going to address this question in detail in  a separate publication.

Finally, all conjectured so far Grassmannian integral formulations for form factors (the one in the present paper and the one from  \cite{Wilhelm_Grassmannians_Integrability}) are given by a linear combinations of top-cell like Grassmannian integrals which are, at least in some cases, not manifestly cyclically invariant with respect to permutations of external states (particles) (the corresponding sums of residues for such Grassmannian integrals are cyclically invariant with respect to such permutations). One may wonder whether it is possible to construct a representation for form factors which will be given by a single term and be manifestly cyclically invariant? Also,
it is interesting to find an analogs of the objects considered here within context of twistor string theories (correlation functions of vertex operators corresponding to open string states together with one vertex operator corresponding to closed string state).

\section{Conclusion}\label{p8}
In this article we considered form factors of operators from $\mathcal{N}=4$ SYM stress tensor operator supermultiplet in the special limit of light-like momentum $q^2=0$ carried by operator. For this special case we have conjectured the Grassmannian integral representation valid both for tree-level form factors and for leading singularities of their loop counterparts. The derivation presented is based on the idea, that the Grassmannian integrals for form factors should be regulated with respect to the soft limit of momentum carried by operator compared to the Grassmannian integrals for amplitudes.

We have successfully verified our conjecture by reproducing known results for $\mbox{MHV}_n$, $~\mbox{N}^{k-2}\mbox{MHV}_{k+1}$ and $\mbox{NMHV}_{5}$ form factors as well as correct soft limit with respect to momentum carried by operator. Using the obtained Grassmannian integral representation we have also discussed, on a particular example of $\mbox{NMHV}_{5}$ form factor, the relations between different BCFW representations and  cancellation of spurious poles. It turns out, that everything works very similar to the case of amplitudes.

We hope that the construction and ideas presented here will be useful for further studies of integrability of form factor both at tree and loop level, construction of form factors of more
general operators as well as  for further investigation of relations between $\mathcal{N}=4$ SYM and twistor string theories.

\section*{Acknowledgements}

The authors would like to thank D.I.Kazakov
for valuable and stimulating discussions. L.V. would like to thank Yu-tin Huang for  stimulating discussion and A.A.Zhukov for help with the preparation of manuscript.
This work was supported by RSF grant \#16-12-10306.

\appendix
\section{Form factors of operators from stress tensor operator supermultiplet in $\mathcal{N}=4$ SYM}\label{aA}
This appendix serves as an introduction to some essential ideas and notation\footnote{In what follows  we will avoid writing  some of indices explicitly in some expressions where it will not lead to confusion.}
regarding the general structure of form factors of operators from  stress-tensor operator supermultiplet formulated in harmonic superspace.

To describe  stress-tensor operator supermultiplet in a manifestly supersymmetric and $SU(4)_R$ covariant way it is useful to consider  harmonic superspace parameterized by the following set of coordinates \cite{N=4_Harmonic_SS,SuperCor1}:
\begin{eqnarray}
\mbox{$\mathcal{N}=4$ harmonic
superspace}&=&\{x^{\alpha\dot{\alpha}},
~\theta^{+a}_{\alpha},\theta^{-a'}_{\alpha},
~\bar{\theta}_{a~\dot{\alpha}}^{+},\bar{\theta}_{a'~\dot{\alpha}}^{-}, u_A^{+a},u_A^{-a'}
\}.
\end{eqnarray}
Here $u_A^{+a}$, $u_A^{-a'}$ is a set of harmonic coordinates parameterizing coset
$$
\frac{SU(4)}{SU(2) \times SU(2)' \times U(1)} ,
$$
$A$ is $SU(4)$ index, $a$ and $a'$ are $SU(2)$ indices, $\pm$ denote $U(1)$ charges; $\theta$'s are Grassmann coordinates, while $\alpha$ and $\dot{\alpha}$ are  $SL(2,\mathbb{C})$ indices.

The stress-tensor operator supermultiplet is given by
\begin{eqnarray}
T=Tr(W^{++}W^{++})
\end{eqnarray}
where $W^{++}(x,\theta^{+},\bar{\theta}^{+})$ is the harmonic
superfield containing all component fields of $\mathcal{N}=4$ SYM
supermultiplet. The latter are given by six scalars $\phi^{AB}$
(anti-symmetric in the $SU(4)_R$ indices $AB$), four Weyl fermions $\psi^A_{\alpha}$ and  gauge field strength tensor $F^{\mu\nu}$, all transforming in the adjoint representation of $SU(N_c)$ gauge group. We would like to note, that
$W^{++}$ superfield is on-shell in the sense that the algebra of supersymmetry transformations leaving it invariant is closed only if the component fields in $W^{++}$ obey their equations of motion.

Next, to describe on-shell states of $\mathcal{N}=4$ SYM supermultiplet it is convenient to introduce on-shell momentum superspace, which in its harmonic version is given by
\begin{eqnarray}
\mbox{$\mathcal{N}=4$ harmonic
on-shell momentum superspace}&=&\{\lambda_{\alpha},\tilde{\lambda}_{\dot{\alpha}},~\eta^{-}_{a},\eta^{+}_{a'},~u_A^{+a},u_A^{-a'}\}. \nonumber \\
\end{eqnarray}
Here $\lambda_{\alpha},\tilde{\lambda}_{\dot{\alpha}}$ are two commuting $SL(2,\mathbb{C})$ Weyl spinors parameterizing momentum of massless on-shell state $p_{\alpha\dot{\alpha}}=\lambda_{\alpha}\tilde{\lambda}_{\dot{\alpha}}$ ($p^2=0$).
All creation/annihilation operators of on-shell states of $\mathcal{N}=4$ SYM supermultiplet given by two physical polarizations of gluons $|g^-\rangle, |g^+\rangle$, four fermions $|\Gamma^A\rangle$ with positive and four fermions
$|\bar{\Gamma}^A\rangle$ with negative helicity together with six real
scalars $|\phi^{AB}\rangle$ (anti-symmetric in the $SU(4)_R$ indices
$AB$ ) can be combined together into one manifestly supersymmetry invariant
``superstate''
$|\Omega_{i}\rangle=\Omega_{i}|0\rangle$ (index $i$ numerates states)
\begin{eqnarray}\label{superstate}
|\Omega_{i}\rangle=\left(g^+_i + (\eta\Gamma_{i}) +
\frac{1}{2!}(\eta\eta\phi_{i}) +
\frac{1}{3!}(\varepsilon\eta\eta\eta\bar{\Gamma}_i) +
\frac{1}{4!}(\varepsilon\eta\eta\eta\eta )g^-_i\right)
|0\rangle,
\end{eqnarray}
where $(\ldots)$ schematically represents contractions with respect to
the $SU(2) \times SU(2)' \times U(1)$
indices and $(\varepsilon\ldots)$ represents additional contraction
with $\varepsilon_{ABCD}$ symbol. It is assumed that all $SU(4)$ indices should be expressed in terms of $SU(2) \times SU(2)' \times U(1)$ indices  using harmonic variables $u$. The $n$ particle superstate
$|\Omega_1\ldots\Omega_n\rangle$ is then given by
$|\Omega_1\ldots\Omega_n\rangle=\prod_{i=1}^n\Omega_i|0\rangle$.
It turns out that to obtain form factors of full stress tensor operator supermultiplet at tree level it is enough to consider only its chiral
or self dual truncation, which is realized by simply putting all $\bar{\theta}$ to zero in $T$:
\begin{eqnarray}\label{superstate}
\mathcal{T}(x,\theta^+)=Tr(W^{++}W^{++})|_{\bar{\theta}=0}.
\end{eqnarray}
All operators in $\mathcal{T}$ supermultiplet are constructed using the fields from the self dual part of the full $\mathcal{N}=4$ SYM supermultiplet. It is important to note that all component fields in $\mathcal{T}$ may be considered off-shell now. Using on-shell momentum and harmonic $\mathcal{N}=4$ SYM superspaces the functional dependence of color ordered form factors $Z_n$
of operators from the chiral truncation of stress-tensor operator supermultiplet
could be written as
\begin{eqnarray}
	\langle\Omega_1\ldots\Omega_n|\mathcal{T}(q,\gamma^{-})|0\rangle=Z_n(\{\lambda,\tilde{\lambda},\eta\};q,\gamma^{-}),
\end{eqnarray}
where $\{\lambda,\tilde{\lambda},\eta\}$ are parameters of the external on-shell states, while $\gamma^{-}$ and $q$ parametrize the operator content of the chiral part of $\mathcal{N}=4$ SYM stress-tensor operator supermultiplet and its momentum. It is assumed that the following  transformation from $x,\theta^{+}$ to $q,\gamma^{-}$ was performed
\begin{equation}
\hat{T}[\ldots] = \int d^4x^{\alpha\dot{\alpha}}~d^{-4}\theta
\exp(iqx+\theta^{+}\gamma^{-})[\ldots].
\end{equation}

Using invariance under supersymmetry transformations ($Z_n$ should be annihilated
by an appropriate set of supercharges) we can further fix the Grassmann structure of the form factor (see \cite{HarmonyofFF_Brandhuber,BKV_SuperForm} for more detais):
\begin{eqnarray}\label{T[superFormfactor]}
Z_n (\{\lambda,\tilde{\lambda},\eta\};q,\gamma^{-}) &=&
\delta^4(\sum_{i=1}^n\lambda_{\alpha}^i\tilde{\lambda}_{\dot{\alpha}}^i-q_{\alpha\dot{\alpha}})
\delta^{-4}(q^-_{a\alpha}+\gamma^-_{a\alpha})\delta^{+4}(q^+_{a'\alpha})
\mathcal{X}_n\left(\{\lambda,\tilde{\lambda},\eta\}\right),\nonumber\\
\mathcal{X}_n&=&\mathcal{X}_n^{(0)} + \mathcal{X}_n^{(4)} + \ldots +
\mathcal{X}_n^{(4n-8)}
\end{eqnarray}
and
\begin{eqnarray}
q^{+}_{a'\alpha}=\sum_{i=1}^n\lambda_{\alpha}^i\eta^{+}_{a'i},
~q^{-}_{a\alpha}=\sum_{i=1}^n\lambda_{\alpha}^i\eta^{-}_{ai},
\end{eqnarray}
Here $\mathcal{X}^{(4m)}_n$ are the homogeneous $SU(4)_R$ and $SU(2)\times SU(2)' \times U(1)$ invariant polynomials of the order $4m$ in Grassmann variables. The structure (\ref{T[superFormfactor]}) is valid both at tree and loop level. The Grassmann $\delta$-functions which one could encounter in this article are given by:
\begin{eqnarray}
\delta^{-4}(q_{a\alpha}^-)=\sum_{i,j=1}^n\prod_{a,b=1}^2\langle ij \rangle \eta^{-}_{a,i}\eta^{-}_{b,j},~~
\delta^{+4}(q_{a\alpha}^+)=\sum_{i,j=1}^n\prod_{a',b'=1}^2\langle ij \rangle
\eta^{+}_{a',i}\eta^{+}_{b',j},
\end{eqnarray}
and
\begin{eqnarray}
\hat{\delta}^{-2}(X^{-a})=\prod_{a=1}^2X^{-a},~~\hat{\delta}^{+2}(X^{+}_{a'})=\prod_{a=1}^2X^{+}_{a'}.
\end{eqnarray}
For convenience we have also introduced the following shorthand notations for bosonic and Grassmann $\hat{\delta}^{4}$ delta-functions:
\begin{eqnarray}
\delta^{-4}\delta^{+4}\equiv\delta^{8},~\hat{\delta}^{-2}\hat{\delta}^{+2}\equiv\hat{\delta}^4,~
\delta^{8}(q+\gamma)\equiv\delta^{8}(q_{1...n}+\gamma)\equiv\delta^{-4}(q_{a\alpha}^-+\gamma_{a\alpha}^-)\delta^{+4}(q_{a'\alpha}^+) ,
\end{eqnarray}
and
\begin{eqnarray}
\hat{\delta}^4(ijk)\equiv \hat{\delta}^4(\eta_i[jk]+\eta_j[ki]+\eta_k[ij]).
\end{eqnarray}
The strings of spinor products were abbreviated as
\begin{eqnarray}
\mathcal{P}(1\ldots n)\equiv\langle 12\rangle\langle23\rangle...\langle n1\rangle,~
\mathcal{P}^*(1\ldots n)\equiv[12][23]...[n1].
\end{eqnarray}

Note that the condition $q^2=0$ doesn't change much in the general structure (\ref{T[superFormfactor]}) of form factor. The condition $q^2=0$ just allows us
to decompose operator momentum as $q_{\alpha\dot{\alpha}}=\lambda_{\alpha,q}\tilde{\lambda}_{\dot{\alpha},q}$. Using
momentum conservation we may always get rid of $q$
dependence in $\mathcal{X}_n$, which we emphasized by writing $\mathcal{X}_n\left(\{\lambda,\tilde{\lambda},\eta\}\right)$.
On the other hand it is not necessary and we actually find more convenient to keep $q$ dependence in $\mathcal{X}_n$'s in the present paper.

It is easy to see, that $\mathcal{X}^{(0)}_n$, $\mathcal{X}^{(4)}_n$ etc. are
analogs of MHV, NMHV etc.
parts of the superamplitude \cite{DualConfInvForAmplitudesCorch}. For example, the part of super form factor proportional to $\mathcal{X}^{(0)}_n$ contains component form factors with the overall helicity of external states equal to $n-2$. The latter are known as MHV form factors.  Part of super form factor proportional to $\mathcal{X}^{(4)}_n$ contains component form factors with overall helicity $n-4$, so called NMHV form factors  and so on up to $\mathcal{X}_n^{(4n-8)}$ $\overline{\mbox{MHV}}$ form factors with overall helicity $2-n$.

In \cite{HarmonyofFF_Brandhuber} it was claimed that at least at tree level
it is still possible to describe the form factors of the full non-chiral stress tensor operator supermultiplet using full $W^{++}(x,\theta^+,\bar{\theta}^+)$
superfields. All the essential information is  contained in $\mathcal{X}_n\left(\{\lambda,\tilde{\lambda},\eta\}\right)$
functions, which could be computed in the chiral truncated sector and
the form factors of the full stress tensor operator supermultiplet could then be recovered from them. Introducing  non-chiral on-shell momentum superspace together
with Grassmann Fourier transform from $\eta^+_i$ to $\bar{\eta}^-_i$ variables and performing $\hat{T}$ transformation from $(x,\theta^{+},\bar{\theta}^+)$ to $(q,\gamma^-,\bar{\gamma}^-)$ with account for supersymmetry constraints the form factors of the full stress tensor operator supermultiplet $Z_n^{full} $ could be written as
\begin{eqnarray}\label{T[superFormfactorfull]}
Z_n^{full} (\{\lambda,\tilde{\lambda},\eta,\bar{\eta}\},\{q,\gamma^{-},\bar{\gamma}^{-}\}) &=&
\delta^4(\sum_{i=1}^n\lambda_{\alpha}^i\tilde{\lambda}_{\dot{\alpha}}^i-q_{\alpha\dot{\alpha}})
\delta^{-4}(q^-_{a\alpha}+\gamma^-_{a\alpha})
\delta^{-4}(\bar{q}^{-a'}_{\alpha}+\bar{\gamma}^{-a'}_{\alpha})\times\nonumber\\
&\times&\int \prod_{k=1}^n d^{+2}\eta_k~exp(\eta^+_k\bar{\eta}^-_k)
\delta^{+4}(q^+_{a'\alpha})
\mathcal{X}_n\left(\{\lambda,\tilde{\lambda},\eta\}\right),\nonumber\\
\end{eqnarray}
In the present article however we will work only with the chiral truncation of stress-tensor operator supermultiplet.

Using the BCFW recursion relations \cite{FormFactorMHV_component_Brandhuber}
one can show that MHV form factors could be written as (here we dropped the momentum conservation $\delta$-function)
\begin{eqnarray}
Z_n^{(2)}=\delta^{8}(q+\gamma)\mathcal{X}^{(0)}_n,~\mathcal{X}^{(0)}_n=\frac{1}{\mathcal{P}(1\ldots n)}.
\end{eqnarray}
It is instructive to compare them with well known results for the tree level
$\mbox{MHV}_n$ and $\overline{\mbox{MHV}}_3$ amplitudes  given by
\begin{equation}
A_n^{(2)}=\frac{\delta^{8}(q)}{\mathcal{P}(1\ldots n)},~
A_3^{(1)}=\frac{\hat{\delta}^{4}(\eta_1[23]+\eta_2[31]+\eta_3[12])}{\mathcal{P}^*(123)}.
\end{equation}

Finally, let us comment on the value of numerical coefficient in the relation (\ref{cojectureAmpl-FF}). Schematically BCFW recursion for $Z^{(k)}_n$ form factors could be written as
\begin{eqnarray}\label{BCFWTwistorSchem}
Z^{(k)}_n=\sum A_{n_1}^{L,k_1}\otimes Z_{n_2}^{R,k_2}+
\sum Z_{n_1}^{L,k_1}\otimes A_{n_2}^{R,k_2},
\end{eqnarray}
where symbol $\otimes$ denotes corresponding BCFW shifts and in the sums above it is understood that $k_1+k_2=k+1$ and $n_1+n_2 = n+2$. An explicit analysis of BCFW diagrams together with analytical expression for $Z^{(2)}_n$ form factors shows  that $Z^{(k)}_n\big{|}_{q,\gamma=0}=C^kA^{(k)}_n$, for $k=2,3$ with $C^2=1$, $C^3=2$ (see Figs. \ref{BCFWdiagramsNMHV4} and \ref{BCFWdiagramsNMHV5} as an example). Proceeding by induction and taking a corresponding limit for (\ref{BCFWTwistorSchem}) we have
\begin{eqnarray}\label{}
Z^{(k)}_n\big{|}_{q,\gamma=0}=\sum\left(C^{k_1}+C^{k_2}\right) A_{n_1}^{L,k_1}\otimes A_{n_2}^{R,k_2}.
\end{eqnarray}
Self consistency of BCFW recursion then requires that $C^k=\left(C^{k_1}+C^{k_2}\right)$,
which can be easily solved and we get $C^k=k-1$.

\section{Evaluation of $\mbox{NMHV}_{4,5}$,   $\mbox{N}^2\mbox{MHV}_{5}$ and $\mbox{N}^3\mbox{MHV}_{6}$ form factors via Grassmannian integral}\label{aB}
In this appendix we decided to give more details on the evaluation of form factors presented in the main body of the paper both using BCFW  recursion and Grassmannian integral representation. Let us start with $\mbox{NMHV}_n$ sector first. BCFW recursion can be solved in this case and we get
\begin{eqnarray}\label{SolutionOfBCFWNMHVsector}
    Z_{n}^{(3)}=Z_{n}^{(2)}\left(\sum_{i=2}^{n-2}\sum_{j=i+1}^{n-1}R^{(1)}_{1ji}+
    \sum_{i=2}^{n-2}\sum_{j=i+2}^{n}R^{(2)}_{1ji}\right),
\end{eqnarray}
where $R_{rst}^{(1,2)}$ functions are defined as
\cite{BORK_NMHV_FF}:
\begin{figure}[t]
    \begin{center}
        \epsfxsize=6cm
        \epsffile{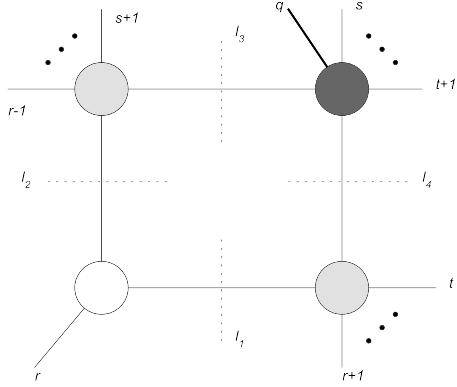}
    \end{center}\vspace{-0.2cm}
    \caption{Diagrammatic representation of the
        quadruple cut proportional to $R^{(1)}_{rst}$.
        The dark grey blob is the MHV form factor.}\label{NMHV_R1}
\end{figure}
\begin{figure}[t]
    \begin{center}
        \epsfxsize=6cm
        \epsffile{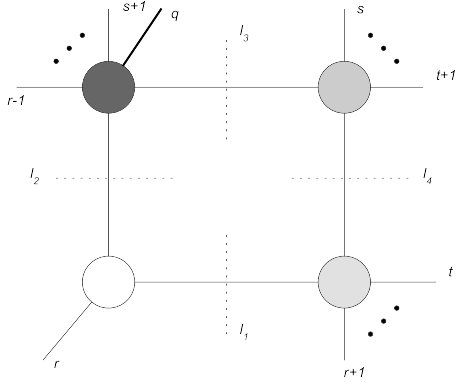}
    \end{center}\vspace{-0.2cm}
    \caption{Diagrammatic representation of the quadruple cut proportional to $R^{(2)}_{rst}$. }\label{NMHV_R2}
\end{figure}
\begin{eqnarray}\label{R_1_deff}
    R_{rst}^{(1)}&=&\frac{\langle s+1s\rangle\langle
        t+1t\rangle\hat{\delta}^4\left(\sum_{i=t}^{r+1}\eta_i\langle
        i|p_{t...s+1}p_{r...s+1}|r\rangle-\sum_{i=r}^{s+1}\eta_i\langle
        i|p_{t...s+1}p_{t...r+1}|r\rangle\right)} {p_{s+1...t}^2\langle
        r|p_{r...s+1}p_{t...s+1}|t+1\rangle\langle
        r|p_{r...s+1}p_{t...s+1}|t\rangle\langle
        r|p_{t...r+1}p_{t...s+1}|s+1\rangle\langle
        r|p_{t...r+1}p_{t...s+1}|s\rangle},\nonumber\\
\end{eqnarray}
\begin{eqnarray}\label{R_2_deff}
    R_{rst}^{(2)}&=&\frac{\langle s+1s\rangle\langle
        t+1t\rangle\hat{\delta}^4\left(\sum_{i=t}^{r+1}\eta_i\langle
        i|p_{s...t+1}p_{s...r+1}|r\rangle-\sum_{i=r+1}^{s}\eta_i\langle
        i|p_{s...t+1}p_{t...r+1}|r\rangle\right)} {p_{s...t+1}^2\langle
        r|p_{s...r+1}p_{s...t+1}|t+1\rangle\langle
        r|p_{s...r+1}p_{s...t+1}|t\rangle\langle
        r|p_{t...r+1}p_{s...t+1}|s+1\rangle\langle
        r|p_{t...r+1}p_{s...t+1}|s\rangle}.\nonumber\\
\end{eqnarray}
In the case of $\mbox{NMHV}_4$ form factor we have
\begin{eqnarray}\label{ZNMHV4}
Z^{(3)}_{4}=Z^{(2)}_{4}\left(R^{(1)}_{132}+R^{(2)}_{142}\right),
\end{eqnarray}
where after some simplifications
\begin{eqnarray}\label{ZNMHV4full}
Z^{(2)}_{4}R^{(1)}_{132}&=&\delta^8(q_{1\ldots4}+\gamma)
\frac{\langle3q\rangle\hat{\delta}^4(124)}{\langle q3\rangle^4[12][2q][3q][q4][41]},\nonumber\\
Z^{(2)}_{4}R^{(2)}_{142}&=&\delta^8(q_{1\ldots4}+\gamma)
\frac{\langle1q\rangle\hat{\delta}^4(124)}{\langle 43\rangle^4[1q][2q][23][34][4q]}.
\end{eqnarray}
Combining these terms together we get
\begin{eqnarray}\label{ZNMHV4ZNMHV4simpl}
Z^{(3)}_{4}&=&\delta^8(q_{1\ldots4}+\gamma)\frac{\hat{\delta}^4(123)}{\langle4q\rangle^4}
\left(\frac{\langle1q\rangle[12][q3][14]+
\langle3q\rangle[23][34][1q]}{[1q][2q][3q][4q]\mathcal{P}^*(1234)}\right).
\end{eqnarray}
Grassmannian integral in the case of $\mbox{NMHV}_4$ form factor is over $Gr(3,5)$ Grassmannian and is fully localized on $\delta$ functions. The result of integration is given by (\ref{ZNMHV4ZNMHV4simpl}). This result should be cyclically symmetric with respect to permutations of external states, i.e. with respect to the action of permutation operator $\mathbb{P}$ shifting state numbers by $+1$ and leaving position of $q$ intact. With the use of momentum twistor representation it is easy to see that the combination
\begin{eqnarray}\label{ZNMHV4ZNMHV4simpl}
\delta^8(q_{1\ldots4}+\gamma)\frac{\hat{\delta}^4(123)}{\langle4q\rangle^4},
\end{eqnarray}
is invariant with respect to the action of $\mathbb{P}$ in a sense that, for example,
the  coefficients of $\eta^4_1\eta^4_2\eta^4_3$, $\gamma^-=0$ evaluated from
\begin{eqnarray}\label{ZNMHV4ZNMHV4simpl}
\delta^8(q_{1\ldots4}+\gamma)\frac{\hat{\delta}^4(123)}{\langle4q\rangle^4}~
\mbox{or}~\mathbb{P}\left(\delta^8(q_{1\ldots4}+\gamma)\frac{\hat{\delta}^4(123)}{\langle4q\rangle^4}\right)=\delta^8(q_{1\ldots4}+\gamma)\frac{\hat{\delta}^4(234)}{\langle1q\rangle^4}
\end{eqnarray}
give identical results. The combination
\begin{eqnarray}\label{ZNMHV4ZNMHV4simpl}
[1q][2q][3q][4q]\mathcal{P}^*(1234).
\end{eqnarray}
is manifestly invariant with respect to $\mathbb{P}$. And we verified numerically
\cite{SM} that
\begin{eqnarray}\label{ZNMHV4perm}
\mathbb{P}(\langle1q\rangle[12][q3][14]+\langle3q\rangle[23][34][1q])=
\langle1q\rangle[12][q3][14]+\langle3q\rangle[23][34][1q].
\end{eqnarray}
So, as expected, the results obtained for $Z^{(3)}_{4}$ form factor using both BCFW recursion and Grassmannian integral representation are cyclically invariant.
It would be also interesting to write down (\ref{ZNMHV4perm}) in manifestly cyclically invariant form.

Now lets turn to $Z^{(3)}_{5}$ form factor. From BCFW recursion we get
\begin{eqnarray}\label{ZNMHV5}
Z^{(3)}_{5}&=&Z^{(2)}_{5}\left(R^{(1)}_{132}+R^{(1)}_{142}+R^{(1)}_{153}
+R^{(2)}_{152}+R^{(2)}_{142}+R^{(2)}_{153}\right)\nonumber\\&=&A1+B1+C1+A2+B2+C2,
\end{eqnarray}
where each term can be simplified and written in the following form
\begin{eqnarray}\label{ZNMHV5_1}
A1&=&Z_{5}^{(2)}R^{(2)}_{152}=\frac{\langle1q\rangle\delta^8(q_{1...5}+\gamma)\hat{\delta}^4(12q)}
{\langle34\rangle\langle45\rangle[1q][2q]\langle5|3+4|2]\langle3|4+5|q]p_{345}^2},\\
B1&=&Z_{5}^{(2)}R^{(2)}_{142}=\frac{\delta^8(q_{1...5}+\gamma)\hat{\delta}^4(234)}
{\langle15\rangle[43][23]\langle1|5+q|4]\langle5|4+3|2]p_{234}^2},
\end{eqnarray}
\begin{eqnarray}\label{ZNMHV5_2}
A2&=&Z_{5}^{(2)}R^{(1)}_{132}=\frac{\langle3q\rangle\delta^8(q_{1...5}+\gamma)\hat{\delta}^4(234)}
{\langle45\rangle\langle15\rangle[3q][2q]\langle1|5+4|q]\langle4|5+1|2]p_{154}^2},\\
B2&=&Z_{5}^{(2)}R^{(1)}_{142}=\frac{\langle3q\rangle\delta^8(q_{1...5}+\gamma)\hat{\delta}^4(251)}
{\langle43\rangle\langle3q\rangle[12][15]\langle3|1+2|5]\langle4|5+1|2]p_{152}^2},
\end{eqnarray}
\begin{eqnarray}\label{ZNMHV5_3}
C1&=&Z_{5}^{(2)}R^{(2)}_{153}=\frac{\langle1q\rangle\delta^8(q_{1...5}+\gamma)\hat{\delta}^4(45q)}
{\langle12\rangle\langle23\rangle[45][q5]\langle1|3+2|4]\langle3|5+4|q]\langle1|5+4|q]},\\
C2&=&Z_{5}^{(2)}R^{(1)}_{153}=\frac{\langle4q\rangle\delta^8(q_{1...5}+\gamma)\hat{\delta}^4(45q)}
{\langle12\rangle\langle23\rangle[4q][q5]\langle3|q+4|5]\langle1|2+3|q]p_{123}^2                 }.
\end{eqnarray}
For comparison let us also write down the result for 6 point $\mbox{NMHV}$ amplitude
$A_{6}^{(3)}= A_{6}^{(2)}R_{142}+A_{6}^{(2)}R_{153}+A_{6}^{(2)}R_{152}$, where
\begin{eqnarray}\label{NMHV6}
A_{6}^{(2)}R_{152}&=&\frac{\delta^8(q_{1...6})\hat{\delta}^4(126)}
{\langle34\rangle\langle45\rangle[12][16]\langle5|3+4|2]\langle3|4+5|6]p_{345}^2},\\
A_{6}^{(2)}R_{142}&=&\frac{\delta^8(q_{1...6})\hat{\delta}^4(234)}
{\langle56\rangle\langle16\rangle[43][23]\langle1|5+6|4]\langle5|4+3|2]p_{234}^2},\\
A_{6}^{(2)}R_{153}&=&\frac{\delta^8(q_{1...6})\hat{\delta}^4(456)}
{\langle12\rangle\langle23\rangle[45][65]\langle1|3+2|4]\langle3|5+4|6]p^2_{123}}.
\end{eqnarray}

Now we are going to reproduce this result from Grassmannian integral representation (\ref{NMHV5FormFactorGrassmannIntegral}). To evaluate integral over the Grassmannian we are following the strategy of \cite{ArkaniHamed_DualitySMatrix,Grassmanians-N4SYM-ABJM}. In general, fixing $GL(k)$ gauge so that the first $k$ columns of $C_{al}$ matrix form an identity matrix and solving $\delta$ - function constraints in (\ref{GrassmannianIntegralLambda})
or (\ref{GrassmannIntegralFormFactors}) leads to the following underdetermined system of linear equations
\begin{eqnarray}\label{underdeterminedSystemOfEquations}
	c_{ai}\lambda_a&=&-\lambda_i,\nonumber\\
	c_{ai}\tilde{\lambda}_i&=&-\tilde{\lambda}_a,
\end{eqnarray}
where $a=1\ldots k$ and $i=k+1\ldots n$. For other $GL(k)$ gauges the structure
of these equations will be similar, the only difference is the values taken by $a,i$ indexes. The general solution of this system
of equations can be parametrized by $(k-2)(n-k-2)$ complex parameters $\tau_A$:
\begin{eqnarray}\label{GeneralSolutionOfUSoLE1}
	c_{ai}(\tau)=c_{ai}^*+d_{aiA}\tau_A,
\end{eqnarray}
where $d_{aiA}$ are some rational functions of $\lambda,\tilde{\lambda}$'s and $c_{ai}^*$ is some particular solution of (\ref{underdeterminedSystemOfEquations}).
Using this solution the bosonic $\delta$ - functions in (\ref{GrassmannianIntegralLambda}) or (\ref{GrassmannIntegralFormFactors}) (here we are discussing (\ref{GrassmannianIntegralLambda}) for concreteness) could be written as\footnote{The number
of $\delta$ functions in LHS and RHS is the same. In LHS we have $2n$ functions, while in RHS we have $k(n-k)+4-(k-2)(n-k-2)$.}
\begin{eqnarray}\label{FromCtoTau1}
	&&\prod_{a=1}^k
 \delta^{2}\left(\tilde{\lambda}_a+\sum_{i=k+1}^nc_{ai}\tilde{\lambda}_i\right)
 \prod_{i=k+1}^n
 \delta^{2}\left(\lambda_i+\sum_{a=1}^k c_{ai}\lambda_a \right)=\nonumber\\
 &=&\delta^4\left(\sum_{j=1}^n\lambda_j\tilde{\lambda}_j\right)~J(\lambda,\tilde{\lambda})~
\int d^{(k-2)(n-k-2)}\tau_A ~\prod_{a=1}^k\prod_{i=k+1}^n\delta\left(c_{ai}-c_{ai}(\tau)\right),
\end{eqnarray}
where $J(\lambda,\tilde{\lambda})$ is the corresponding Jacobian. The integration $\int \frac{d^{n\times k}C_{al}}{Vol[GL(k)]}$ in (\ref{GrassmannianIntegralLambda}) could be removed using $\delta$ - functions  and the only remaining integration will be over $d^{(k-2)(n-k-2)}\tau_A$ variables.
The minors of $C_{al}$ matrix and  Grassmann $\delta$ - functions in (\ref{GrassmannianIntegralLambda}) are also rewritten in terms of $\tau_A$ variables
using (\ref{GeneralSolutionOfUSoLE1}). The expression under integral sign is then a rational function of  $\tau_A$  and the corresponding integral can be further evaluated with the use of (multidimensional) residue theorem.

In the $Gr(3,6)$ case  it is convenient to choose $GL(3)$ gauge as
\begin{eqnarray}
    C=\left( \begin{array}{cccccc}
        1 & c_{12} & 0 & c_{14} & 0 & c_{16} \\
        0 & c_{32} & 1 & c_{34} & 0 & c_{36} \\
        0 & c_{52} & 0 & c_{54} & 1 & c_{56}\end{array} \right).
\end{eqnarray}
Then the non-trivial coefficients of $C_{al}$ matrix are given by $c_{i'j}$, with $i'=1,3,5$
and $j=2,4,6$ and (\ref{FromCtoTau1}) reduces to (in this case $J(\lambda,\tilde{\lambda})=1$ \cite{ArkaniHamed_DualitySMatrix})
\begin{eqnarray}\label{FromCtoTau}
	&&\prod_{i'=1,3,5}
 \delta^{2}\left(\tilde{\lambda}_{i'}+\sum_{i=2,4,6}^nc_{i'j}\tilde{\lambda}_j\right)
 \prod_{j=2,4,6}
 \delta^{2}\left(\lambda_j+\sum_{i=1,3,5}^k c_{i'j}\lambda_{i'} \right)=\nonumber\\
 &=&\delta^4\left(\sum_{i=1}^6\lambda_i\tilde{\lambda}_i\right)
\int d\tau~\prod_{i'=1,3,5}\prod_{j=2,4,6}^n\delta\left(c_{i'j}-c_{i'j}(\tau)\right),
\end{eqnarray}
with
\begin{eqnarray}\label{GeneralSolutionOfUSoLE}
	c_{i'j}(\tau)=c_{ij'}^*+\epsilon_{i'k'p'}\epsilon_{jlm}\langle k'p'\rangle [lm]~\tau.
\end{eqnarray}
In the case of (\ref{NMHV5FormFactorGrassmannIntegral}) Grassmannian integral $\lambda$'s and $\tilde{\lambda}$'s should be taken from the set
\begin{eqnarray}
(1,2,3,4,5,q)~\mbox{or}~(1,2,3,q,4,5).
\end{eqnarray}
Note also that in the case of form factors we should shift numeration $n\mapsto n+1$ compared to amplitude case. The minors $M_{1},\ldots,M_6$ of $C_{al}$ matrix are linear functions in $\tau$ and corresponding integral over $\tau$ could be evaluated using residues\footnote{We are assuming that the behavior of the particular component extracted from the Grassmann $\delta$ - functions in the numerator of the integrand is no worse then $1/\tau^2$ at infinity. After evaluation of particular residue we supersymmetrize the result assuming that the Grassmann structure should be like $\delta^8(q_{1...5}+\gamma)\hat{\delta}^4(ijk)$.}. To reproduce $Z^{(3)}_{5}$ form factor we should take residues at zeros of $M_1$, $M_3$ and $M_5$ minors. In the chosen gauge these minors are given by
$M_1=c_{52}(\tau)$, $M_3=c_{14}(\tau)$ and $M_5=c_{36}(\tau)$. To simplify the evaluation of residues even further one should note that for each of the residues the
particular solution $c^*_{i'j}$ could be chosen independently such that $c_{52}^*=0$ for $M_1$, $c_{14}^*=0$ for $M_3$, and $c_{36}^*=0$ for $M_5$. Then each residue
corresponds to  $\tau=0$ and all coefficients $c_{i'j}(\tau=0)$ are easily evaluated. For reader's convenience we have gathered the values of all $C_{al}$ matrix elements at poles $1/M_1,\ldots, 1/M_6$ and $1/(135)$ in appendix \ref{aC}. The residues at poles $1/M_1$,$1/M_3$ and $1/M_5$, which we denoted as $\{1\},\{3\}$ and $\{5\}$, of the integral $\Omega^{(3)}_5$
\begin{eqnarray}\label{NMHV5FormFactorGrassmannIntegralAppendix}
\Omega^{(3)}_5=\int \left(\langle1q\rangle \frac{(356)}{(135)}\right)
\frac{\delta^{4|4}(1,2,3,4,5,q)}{M_1\ldots M_6}+\int \left(\langle1q\rangle \frac{(345)}{(135)}+\langle4q\rangle
\frac{(134)}{(135)} \right)\frac{\delta^{4|4}(1,2,3,q,4,5)}{M_1\ldots M_6}
.\nonumber\\
\end{eqnarray}
are given by
\begin{eqnarray}
\{1\}=\tilde{C1}+\tilde{C2},~\{3\}=A1+A2,~\{5\}=B1+B2,
\end{eqnarray}
with
\begin{eqnarray}
\tilde{C1}&=&\langle 1q \rangle\frac{\langle 3|1+2|4]}{\langle 13 \rangle [4q]}
\frac{\delta^8(q_{1...5}+\gamma)\hat{\delta}^4(45q)}
{\langle12\rangle\langle23\rangle[45][q5]\langle1|3+2|4]\langle3|5+4|q]p_{123}^2},\\
\tilde{C2}&=&\langle 3q \rangle\frac{\langle 1|2+3|5]}{\langle 13 \rangle [5q]}\frac{\delta^8(q_{1...5}+\gamma)\hat{\delta}^4(45q)}
{\langle12\rangle\langle23\rangle[4q][45]\langle1|2+3|q]\langle3|q+4|5]p_{123}^2                 }.
\end{eqnarray}
Here, $\tilde{C1}$ term is the result of evaluating the residue at $1/M_1$ pole
in the first integral  while $\tilde{C2}$ is the result of taking the same residue for the second integral. We have checked numerically \cite{SM} that the equality $\tilde{C1}+\tilde{C2}=C1+C2$ holds.
This is the consequence of rather none trivial relations among spinors
($p_1+\ldots+p_5+q=0$ is assumed):
\begin{eqnarray}
&&\frac{\langle1q\rangle\langle3|1+2|4]}{\langle13\rangle[4q][5q]\langle1|2+3|4]\langle3|4+5|1]}
+\frac{\langle3q\rangle\langle1|2+3|5]}{\langle13\rangle[4q][5q]\langle1|2+3|q]\langle3|q+4|5]}=
\nonumber\\
&=&\frac{\langle5q\rangle[45]}{[q4][5q]\langle1|2+3|4]\langle3|4+5|q]}
+\frac{\langle3q\rangle p^2_{123}}{[q4]\langle3|4+5|q]\langle1|2+3|q]\langle3|q+4|5]}.
\end{eqnarray}
So we see that the integral
$\Omega^{(3)}_5$ over contour $\Gamma_{135}$ encircling $M_{1,3,5}$ poles reproduces
BCFW result for $Z^{(3)}_{5}$.

Using (\ref{GrassmannIntegralFormFactors}) one can also easily compute several other cases when
the integral over Grassmannian is fully localized over $\delta$ - functions (that is in the case when $(k-2)(n-k-2)=0$). For $\mbox{N}^2\mbox{MHV}_5$ and $\mbox{N}^3\mbox{MHV}_6$ form factors we have:
\begin{eqnarray}\label{ZNNMHV5}
Z^{(4)}_{5}&=&\delta^8(q_{1\ldots5}+\gamma)\Big{(}
\frac{\hat{\delta}^4(12q)\hat{\delta}^4(345)}{(p^2_{345})^4}
\frac{\langle q|p_1|2]}{[q2]\mathcal{P}^*(12345q)}+
\frac{\hat{\delta}^4(125)\hat{\delta}^4(34q)}{(p^2_{125})^4}
\frac{\langle q|p_1+p_5|2]}{[q2]\mathcal{P}^*(1234q5)}+\nonumber\\
&+&\frac{\hat{\delta}^4(125)\hat{\delta}^4(34q)}{(p^2_{125})^4}
\frac{\langle q|p_3|2]}{[q2]\mathcal{P}^*(123q45)}\Big{)},
\end{eqnarray}
\begin{eqnarray}\label{ZNNNMHV6}
Z^{(5)}_{6}&=&\delta^8(q_{1\ldots6}+\gamma)\Big{(}
\frac{\hat{\delta}^4(12q)\hat{\delta}^4(345)\hat{\delta}^4(62q)}{[2q]^4(p^2_{345})^4}
\frac{\langle q|p_1+p_6|2]}{[q2]\mathcal{P}^*(123456q)}+\nonumber\\
&+&\frac{\hat{\delta}^4(126)\hat{\delta}^4(345)\hat{\delta}^4(6q2)}{[26]^4(p^2_{345})^4}
\frac{\langle q|p_1+p_6|2]}{[q2]\mathcal{P}^*(12345q6)}+
\frac{\hat{\delta}^4(126)\hat{\delta}^4(34q)\hat{\delta}^4(625)}{[26]^4(p^2_{34q})^4}
\frac{\langle q|p_3+p_4|2]}{[q2]\mathcal{P}^*(1234q56)}+\nonumber\\
&+&\frac{\hat{\delta}^4(125)\hat{\delta}^4(34q)}{[16]^4(p^2_{3q4})^4}
\frac{\langle q|p_3|2]}{[q2]\mathcal{P}^*(123q456)}\Big{)}.
\end{eqnarray}
Similar results also gives BCFW recursion for $[1,2\rangle$ shift.

In the case of $\mbox{N}^2\mbox{MHV}_5$ form factor we have also verified numerically \cite{SM}
the cyclical symmetry of particular super form factor  components  corresponding to the form factors of operator given by the Lagrangian of $\mathcal{N}=4$ SYM. Taking gluons as
external states (particles) we have ($\eta^4_1\eta^4_2\eta^4_3\eta^4_4,~\gamma^-=0$):
\begin{eqnarray}\label{ZNNMHV5cyclicalSymm}
\eta^4_1\eta^4_2\eta^4_3\eta^4_4&\sim&[5q]^4\Big{(}
\frac{-\langle q|p_1|2]}{[q2]\mathcal{P}^*(12345q)}+
\frac{\langle q|p_1+p_5|2]}{[q2]\mathcal{P}^*(1234q5)}+\nonumber\\
&+&\frac{\langle q|p_3|2]}{[q2]\mathcal{P}^*(123q45)}\Big{)}.
\end{eqnarray}
This expression is indeed invariant with respect to permutation $\mathbb{P}$ in a sense that
\begin{eqnarray}
Z^{(5)}_{6}\big{|}^{\gamma^-=0}_{\eta^4_1\eta^4_2\eta^4_3\eta^4_4}=
\left(\mathbb{P}Z^{(5)}_{6}\right)\big{|}^{\gamma^-=0}_{\eta^4_1\eta^4_2\eta^4_3\eta^4_4}.
\end{eqnarray}

\section{Residues of $\Omega^{(3)}_5$ }\label{aC}
In this appendix we collected the results for the elements of $C_{al}$ matrix evaluated
at zeroes of minors $M_1,\ldots, M_6$ and $(135)$. Lets start with the first ("left") term in (\ref{NMHV5FormFactorGrassmannIntegralAppendix}) with  $\lambda,\tilde{\lambda}$'s taken from the set $(1,2,3,4,5,q)$. To compute the residues at poles $1/M_{1,3,5}$ the $GL(3)$ gauge was fixed so that the columns $1,3,5$ of $C_{al}$ formed unity matrix and we got:
\begin{eqnarray}
    C\big{|}_{M_1}=\left( \begin{array}{cccccc}
        1 & c_{12}=\dfrac{\langle23\rangle}{\langle13\rangle} & 0 &
        c_{14}=\dfrac{\langle3|1+2|q]}{\langle13\rangle[4q]} & 0 & c_{16}=\dfrac{\langle3|1+2|4]}{\langle13\rangle[4q]} \\
        0 & c_{32}=\dfrac{\langle12\rangle}{\langle13\rangle} & 1 &
        c_{34}=\dfrac{\langle1|2+3|q]}{\langle13\rangle[4q]} & 0 & c_{36}=\dfrac{\langle1|2+3|4]}{\langle13\rangle[4q]} \\
        0 & c_{52}=0 & 0 & c_{54}=\dfrac{[5q]}{[4q]} & 1 & c_{56}=\dfrac{[54]}{[4q]}\end{array} \right),
\end{eqnarray}
\begin{eqnarray}
    C\big{|}_{M_3}=\left( \begin{array}{cccccc}
        1 & c_{12}=\dfrac{[1q]}{[1q]} & 0 & c_{14}=0 & 0 & c_{16}=\dfrac{[12]}{[q2]} \\
        0 & c_{32}=\dfrac{\langle5|3+4|q]}{\langle35\rangle[2q]} & 1 & c_{34}=\dfrac{\langle45\rangle}{\langle35\rangle} & 0 & c_{36}=\dfrac{\langle5|3+4|2]}{\langle 35\rangle[2q]} \\
        0 & c_{52}=\dfrac{\langle3|4+5|q]}{\langle53\rangle[2q]} & 0 & c_{54}=\dfrac{\langle 43\rangle}{\langle 53 \rangle} & 1 & c_{56}=\dfrac{\langle 3|4+5|2]}{\langle53\rangle [q2]}\end{array} \right),
\end{eqnarray}
\begin{eqnarray}
    C\big{|}_{M_5}=\left( \begin{array}{cccccc}
        1 & c_{12}=\dfrac{\langle5|1+q|4]}{\langle15\rangle[24]} & 0 & c_{14}=\dfrac{\langle5|1+q|2]}{\langle15\rangle[24]} & 0 & c_{16}=\dfrac{\langle 5q\rangle}{\langle15\rangle} \\
        0 & c_{32}=\dfrac{[34]}{[24]} & 1 & c_{34}=\dfrac{[32]}{[42]} & 0 & c_{36}=0 \\
        0 & c_{52}=\dfrac{\langle1|5+q|4]}{\langle15\rangle[24]} & 0 & c_{54}=\dfrac{\langle1|5+q|2]}{\langle15\rangle[24]} & 1 & c_{56}=\dfrac{\langle q1\rangle}{\langle 51 \rangle}\end{array} \right).
\end{eqnarray}
For the residues at $1/M_{2,4,6}$ poles the $GL(3)$ gauge was chosen so that the unity matrix was formed by the columns $2,4,6$:
\begin{eqnarray}
    C\big{|}_{M_2}=\left( \begin{array}{cccccc}
        c_{21}=\dfrac{\langle 4|2+3|5]}{\langle24\rangle[51]} & 1 & c_{23}=\dfrac{\langle 34\rangle}{\langle24\rangle} & 0 & c_{25}=\dfrac{\langle4|2+3|1]}{\langle24\rangle[51]} &0  \\
        c_{41}=\dfrac{\langle2|3+4|5]}{\langle 24\rangle[51]} & 0 & c_{43}=\dfrac{\langle32\rangle}{\langle24\rangle} & 1 & c_{45}=\dfrac{\langle2|3+4|1]}{\langle 24\rangle[51]} & 0 \\
        c_{61}=\dfrac{[5q]}{[51]} & 0 & c_{63}=0 & 0 & c_{65}=\dfrac{[1q]}{[15]} & 1\end{array} \right),
\end{eqnarray}
\begin{eqnarray}
    C\big{|}_{M_4}=\left( \begin{array}{cccccc}
        c_{21}=\dfrac{[23]}{[13]} & 1 & c_{23}=\dfrac{[12]}{[13]} & 0 & c_{25}=0 &0  \\
        c_{41}=\dfrac{\langle q|4+5|3]}{\langle 4q \rangle [13]} & 0 & c_{43}=\dfrac{\langle q|4+5|1]}{\langle 4q \rangle[13]} & 1 & c_{45}=\dfrac{\langle 5q\rangle}{\langle 4q\rangle} & 0 \\
        c_{61}=\dfrac{\langle 4| 5+q|3]}{\langle 4q\rangle [13]} & 0 & c_{63}=\dfrac{\langle 4|5+q|1]}{\langle 4q\rangle[13]} & 0 & c_{65}=\dfrac{\langle 45\rangle}{\langle 4q\rangle} & 1\end{array} \right),
\end{eqnarray}
\begin{eqnarray}
    C\big{|}_{M_6}=\left( \begin{array}{cccccc}
        c_{21}=\dfrac{\langle 1q \rangle}{\langle 2q \rangle} & 1 & c_{23}=\dfrac{\langle q|1+2|5]}{\langle 2q \rangle [35]} & 0 & c_{25}=\dfrac{\langle q|1+2|3]}{\langle 2q \rangle [53]} &0  \\
        c_{41}=0 & 0 & c_{43}=\dfrac{[45]}{[35]} & 1 & c_{45}=\dfrac{[43]}{53]} & 0 \\
        c_{61}=\dfrac{\langle 12\rangle}{ \langle q2 \rangle} & 0 & c_{63}=\dfrac{\langle 2|q+1|5]}{\langle q2\rangle[35]} & 0 & c_{65}=\dfrac{\langle 2|1+q|3]}{\langle q2 \rangle [35]} & 1\end{array} \right).
\end{eqnarray}
And finally for the case of residue at $1/(135)$ pole the unit matrix fixing $GL(3)$ symmetry is formed by $1,2,3$ columns of $C_{al}$ matrix
\begin{eqnarray}
    C\big{|}_{(135)}=\left( \begin{array}{cccccc}
        1 & 0 & 0 & c_{14}=\dfrac{\langle 3 |1+5|q]}{\langle 13 \rangle [q4]}& c_{15}=\dfrac{\langle 53\rangle}{\langle 13 \rangle} & c_{16}=\dfrac{\langle 3|1+5|4]}{\langle 31\rangle [q4]}  \\
        0 & 1 & 0 & c_{24}=\dfrac{[2q]}{[4q]} & c_{25}=0 & c_{26}=\dfrac{[24]}{[q4]} \\
        0 & 0 & 1 & c_{34}=\dfrac{\langle 1|3+5|q]}{[4q]\langle 13\rangle} & c_{35}=\dfrac{\langle 15\rangle}{\langle13\rangle} & c_{36}=\dfrac{\langle1|3+5|4]}{\langle13\rangle[q4]}\end{array} \right).
\end{eqnarray}
The results for the second ("right") term in (\ref{NMHV5FormFactorGrassmannIntegralAppendix})) when $\lambda,\tilde{\lambda}$'s are taken from the set $(1,2,3,q,4,5)$  could be obtained from the above expressions with  simple relabeling
\begin{eqnarray}
\begin{matrix}
1&2&3&4&5&q\\
\downarrow&\downarrow&\downarrow&\downarrow&\downarrow&\downarrow\\
1&2&3&q&4&5
\end{matrix}
\end{eqnarray}
At the end, to emphasize the analytical structure of each contribution let us also write down the denominators of each residue. The corresponding expressions for "left" and "right" terms  for $\{1\}$, $\{3\}$, $\{5\}$ are given by
\begin{eqnarray}
&&\{1\}^{L}\sim\frac{1}{\langle12\rangle\langle23\rangle[45][5q]\langle1|2+3|4]\langle3|4+5|q]p^2_{q45}},\nonumber\\
&&\{1\}^{R}\sim\frac{1}{\langle12\rangle\langle23\rangle[45][4q]\langle1|2+3|q]\langle3|4+q|5]p^2_{q45}},
\end{eqnarray}
\begin{eqnarray}
&&\{3\}^{L}\sim\frac{1}{\langle34\rangle\langle45\rangle[1q][2q]\langle5|3+4|2]\langle3|4+5|q]p^2_{12q}},\nonumber\\
&&\{3\}^{R}\sim\frac{1}{\langle45\rangle\langle51\rangle[3q][2q]\langle1|5+4|q]\langle4|5+1|2]p^2_{23q}},
\end{eqnarray}
\begin{eqnarray}
&&\{5\}^{L}\sim\frac{1}{\langle15\rangle[43][23]\langle1|2+3|4]\langle5|4+3|2]p^2_{234}},\nonumber\\
&&\{5\}^{R}\sim\frac{1}{\langle43\rangle[12][15]\langle3|1+2|5]\langle4|5+1|2]p^2_{125}},
\end{eqnarray}
while for $\{2\}$,$\{4\}$,$\{6\}$ and $(135)$ they are given by
\begin{eqnarray}
&&\{2\}^{L}\sim\frac{1}{\langle23\rangle\langle34\rangle[5q][q1]
\langle3|2+4|q]\langle2|3+4|5]\langle4|2+3|1]p^2_{234}},\nonumber\\
&&\{2\}^{R}\sim\frac{1}{\langle23\rangle\langle3q\rangle[45][51]
\langle3|q+2|5]\langle2|3+q|4]\langle q|2+3|1]p^2_{145}},
\end{eqnarray}
\begin{eqnarray}
&&\{4\}^{L}\sim\frac{1}{\langle54\rangle\langle5q\rangle[12][23]
\langle5|1+3|2]\langle4|5+q|1]\langle q|4+5|3]p^2_{123}},\nonumber\\
&&\{4\}^{R}\sim\frac{1}{\langle4q\rangle\langle45\rangle[12][23]
\langle4|1+3|2]\langle q|4+5|1]\langle 5|4+q|3]p^2_{123}},
\end{eqnarray}
\begin{eqnarray}
&&\{6\}^{L}\sim\frac{1}{\langle q1\rangle\langle12\rangle[34][45]\langle1|3+5|4]\langle q|1+2|3]\langle2|1+q|5]p^2_{q12}},\nonumber\\
&&\{6\}^{R}\sim\frac{1}{\langle 51\rangle\langle12\rangle[q3][q4]
\langle1|3+4|q]\langle 5|1+2|3]\langle2|1+5|4]p^2_{512}},
\end{eqnarray}
\begin{eqnarray}
&&\{(135)\}^{L}\sim\frac{1}{\langle15\rangle[2q]
\langle5|1+3|2]\langle1|3+5|4]\langle 3|1+5|q]},\nonumber\\
&&\{(135)\}^{R}\sim\frac{1}{\langle43\rangle[2q]
\langle3|1+4|5]\langle4|1+3|2]\langle 1|3+4|q]}.
\end{eqnarray}
From these expressions we see that spurious poles indeed cancel in the sums of residues for contours $\Gamma_{135}$ and $\Gamma_{246*}$ and come in pairs as needed.


\begin{thebibliography}{99}
	

\bibitem{Reviews_Ampl_General}
Z.~Bern, L.~J.~Dixon, D.~A.~Kosower
\emph{Progress in One-Loop QCD Computations},
Ann.\ Rev.\ Nucl.\ Part.\ Sci.\ \textbf{46} (1996) 109,
arXiv:hep-ph/9602280 v1.\\
Z.~Bern, L.~J.~Dixon, D.A.~Kosower
\emph{On-Shell Methods in Perturbative QCD},
Annal. of Phys. \textbf{322} (2007) 1587, arXiv:0704.2798 [hep-ph],\\
R.~Britto
\emph{Loop amplitudes in gauge theories: modern analytic approaches},
J.\ Phys.\ A \textbf{44}, 454006 (2011), arXiv:1012.4493 v2 [hep-th], \\
Z.~Bern, Yu-tin ~Huang
\emph{Basics of Generalized Unitarity},
J.\ Phys.\ A \textbf{44} (2011) 454003, arXiv:1103.1869 v1 [hep-th].

\bibitem{Henrietta_Amplitudes}
H.~Elvang, Yu-tin Huang, \emph{Scattering Amplitudes},
arXiv:1308.1697 v1 [hep-th].

\bibitem{Nair}
V.~P.~Nair,
\emph{A Current Algebra for Some Gauge Theory Amplitudes},
Phys.\ Lett.\ B {\bf 214}, 215 (1988).

\bibitem{DualConfInvForAmplitudesCorch}
J.~M.~Drummond, J.~Henn, G.~P.~Korchemsky, E.~Sokatchev, \emph{Dual
	superconformal symmetry of scattering amplitudes in $\mathcal{N}=4$
	super-Yang-Mills theory}, Nucl.\ Phys.\  B {\bf 828} (2010) 317,
arXiv:0807.1095 [hep-th].

\bibitem{ArkaniHamed_DualitySMatrix}
N.~Arkani-Hamed, F.~Cachazo, C.~Cheung and J.~Kaplan,
{\it A Duality For The S Matrix},
JHEP {\bf 1003}, 020 (2010),
arXiv:0907.5418 [hep-th].

\bibitem{Arcani_Hamed_PositiveGrassmannians}
N.~Arkani-Hamed, J.~Bourjaily, F.~Cachazo, A.~Goncharov,
A.~Postnikov and J.~Trnka, \emph{Scattering Amplitudes and the
Positive Grassmannian}, (2012), arXiv:1212.5605 v1 [hep-th].

\bibitem{ArkaniHamed_UnificationResidues}
N.~Arkani-Hamed, J.~Bourjaily, F.~Cachazo and J.~Trnka,
{\it Unification of Residues and Grassmannian Dualities},
JHEP {\bf 1101}, 049 (2011),
arXiv:0912.4912 [hep-th].

\bibitem{Hoges_Polytopes}
A.~Hodges, \emph{Eliminating spurious poles from gauge-theoretic amplitudes },
JHEP {\bf 1305} (2013) 135, arXiv:0905.1473 [hep-th].

\bibitem{Masson_Skiner_Grassmaians_Twistors}
L.~J.~Mason, D.~Skinner,
\emph{Dual Superconformal Invariance, Momentum Twistors and Grassmannians},
JHEP {\bf 0911} (2009) 045, arXiv:0909.0250 [hep-th].

\bibitem{Arcani_Hamed_Polytopes}
N.~Arkani-Hamed, J.~L.~Bourjaily, F.~Cachazo, A.~Hodges, J.~Trnka,
\emph{A Note on Polytopes for Scattering Amplitudes},
JHEP {\bf 1204} (2012) 081, arXiv:1012.6030 [hep-th].

\bibitem{Amplituhdron_1}
N.~Arkani-Hamed, J.~Trnka, \emph{The Amplituhedron},
arXiv:1312.2007 [hep-th].

\bibitem{Amplituhdron_2}
N.~Arkani-Hamed, J.~Trnka, \emph{Into the Amplituhedron},
arXiv:arXiv:1312.7878 [hep-th].


\bibitem{Amplituhdron_3}
Y.~Bai and S.~He,
\emph{The Amplituhedron from Momentum Twistor Diagrams},
JHEP {\bf 1502}, 065 (2015),
arXiv:1408.2459 [hep-th].

\bibitem{Amplituhdron_4}
S.~Franco, D.~Galloni, A.~Mariotti and J.~Trnka,
\emph{Anatomy of the Amplituhedron},
JHEP {\bf 1503}, 128 (2015),
arXiv:1408.3410 [hep-th].

\bibitem{Amplituhdron_5}
Z.~Bern, E.~Herrmann, S.~Litsey, J.~Stankowicz and J.~Trnka,
\emph{Evidence for a Nonplanar Amplituhedron},
JHEP {\bf 1606}, 098 (2016), arXiv:1512.08591 [hep-th].

\bibitem{Amplituhdron_6}
L.~Ferro, T.~Lukowski, A.~Orta, M.~Parisi,
\emph{Towards the Amplituhedron Volume},
JHEP {\bf 1603}, 014 (2016),
arXiv:1512.04954 [hep-th].

\bibitem{BeisertYangianRev}
N.~Beisert, \emph{On Yangian Symmetry in Planar $\mathcal{N}=4$ SYM},
arXiv:1004.5423v2 [hep-th].

\bibitem{Staudacher_SpectralReg_New}
N.~Kanning, T.~Lukowski, M.~Staudacher,
\emph{A Shortcut to General Tree-level Scattering Amplitudes in N=4 SYM via Integrability},
Fortsch.Phys. {\bf 62} (2014) 556-572, arXiv:1403.3382 [hep-th].

\bibitem{Beisert_SpectralReg_New}
N.~Beisert, J.~Broedel, M.~Rosso,
\emph{On Yangian-invariant regularisation of deformed on-shell diagrams in N=4 super-Yang-Mills theory},
arXiv:1401.7274 [hep-th].

\bibitem{Derkachev_SpectralReg_New}
D.~Chicherin, S.~Derkachov, R.~Kirschner,
\emph{Yang-Baxter operators and scattering amplitudes in N=4 super-Yang-Mills theory},
Nucl.\ Phys.\ B \textbf{881} (2014) 467-501, arXiv:1309.5748 [hep-th].

\bibitem{deleeuwm_2014_1}
J.~Broedel, M.~de Leeuw, M.~Rosso,
\emph{A dictionary between R-operators, on-shell graphs and Yangian algebras},
JHEP {\bf 1406} (2014) 170, arXiv:1403.3670 [hep-th].

\bibitem{deleeuwm_2014_2}
J.~Broedel, M.~de Leeuw, M.~Rosso,
\emph{Deformed one-loop amplitudes in N = 4 super-Yang-Mills theory},
arXiv:1406.4024 [hep-th].

\bibitem{Frassek_BetheAnsatzYangianInvariants}
R.~Frassek, N.~Kanning, Y.~Ko and M.~Staudacher,
{\it Bethe Ansatz for Yangian Invariants: Towards Super Yang-Mills Scattering Amplitudes},
Nucl.\ Phys.\ B {\bf 883}, 373 (2014)
[arXiv:1312.1693 [math-ph].

\bibitem{FF_in_integrable_sys}
T.~Klose, T.~McLoughlin, \emph{Worldsheet Form Factors in AdS/CFT},
Phys.\ Rev.\ D {\bf  87} (2013) 026004, [arXiv:1208.2020].

\bibitem{HuotEquation}
S.~Caron-Huot, Song He, \emph{Jumpstarting the All-Loop S-Matrix of
Planar N=4 Super Yang-Mills}, arXiv:1112.1060 [hep-th].

\bibitem{Twistors_DescentEquation}
M.~Bullimore, D.~Skinner, \emph{Descent Equations for Superamplitudes},
arXiv:1112.1056 [hep-th].

\bibitem{vanNeerven_InfraredBehaviorFormFactorsN4SYM}
W.~L.~van Neerven,
{\it Infrared Behavior of On-shell Form-factors in a $N=4$ Supersymmetric {Yang-Mills} Field Theory},
Z.\ Phys.\ C {\bf 30}, 595 (1986).

\bibitem{Perturbiner}
K.~G. Selivanov, \emph{On tree form-factors in (supersymmetric)Yang-Mills theory},
Commun. Math. Phys. {\bf 208} (2000) 671,
arXiv:9809046 [hep-th].

\bibitem{FormFactorMHV_component_Brandhuber}
A.~Brandhuber, B.~Spence, G.~Travaglini and G.~Yang,
\emph{Form Factors in $\mathcal{N}=4$ Super Yang-Mills and Periodic Wilson Loops},
JHEP {\bf 1101} (2011) 134, arXiv:1011.1899 [hep-th].

\bibitem{BKV_Form_Factors_N=4SYM}
L.~V.~Bork, D.~I.~Kazakov, G.~S.~Vartanov, \emph{On form factors in $\mathcal{N}=4$ SYM},
JHEP {\bf 1102} (2011) 063, arXiv:1011.2440 [hep-th].

\bibitem{HarmonyofFF_Brandhuber}
A.~Brandhuber, O.~Gurdogan, R.~Mooney, G.~Travaglini, Gang Yang,
\emph{Harmony of Super Form Factors}, JHEP {\bf 1110} (2011) 046,
arXiv:1107.5067 [hep-th].

\bibitem{BKV_SuperForm}
L.~V.~Bork, D.~I.~Kazakov, G.~S.~Vartanov,
\emph{On MHV Form Factors in Superspace for $\mathcal{N}=4$ SYM Theory},
JHEP {\bf 1110} (2011) 133, arXiv:1107.5551 [hep-th].

\bibitem{BORK_NMHV_FF}
L.~V.~Bork, \emph{On NMHV Form Factors in $\mathcal{N}=4$ SYM Theory from generalized unitarity},
JHEP {\bf 01} (2013) 049, arXiv:1203.2596 [hep-th].

\bibitem{FF_MHV_3_2loop}
A.~Brandhuber, G.~Travaglini, Gang Yang, \emph{Analytic two-loop form factors in N=4 SYM},
arXiv:1201.4170 [hep-th].

\bibitem{Zhiboedov_Strong_coupling_FF}
J.~Maldacena and A.~Zhiboedov, \emph{Form factors at strong coupling via a $Y$-system},
JHEP {\bf 1011} (2010) 104, arXiv:1009.1139
[hep-th].

\bibitem{Strong_coupling_FF_Yang_Gao}
Zhiquan Gao, Gang Yang,
\emph{Y-system for form factors at strong coupling in $AdS_5$ and with multi-operator insertions in $AdS_3$},
JHEP {\bf 1306} (2013) 105, arXiv:1303.2668 [hep-th].

\bibitem{BORK_POLY}
L.~V.~Bork, \emph{On Form Factors in $\mathcal{N}=4$ SYM Theory and polytopes},
JHEP {\bf 1412} (2014) 111, arXiv:1407.5568 [hep-th].

\bibitem{Roiban_FormFactorsOfMultipleOperators}
Oluf~Tang~Engelund, R.~Roiban, \emph{Correlation functions of local composite operators from generalized unitarity}, JHEP 1303 (2013) 172, arXiv:1209.0227 [hep-th].

\bibitem{FormFactorMHV_half_BPS_Brandhuber}
B.~Penante, B.~Spence,~G.~Travaglini, C.~Wen,
\emph{On super form factors of half-BPS operators in $\mathcal{N}=4$ SYM.},
arXiv:1402.1300 [hep-th].

\bibitem{FormFactorMHV_Remainder_half_BPS_Brandhuber}
A.~Brandhuber, B.~Penante, G.~Travaglini, C.~Wen,
\emph{The last of the simple remainders}, arXiv:1406.1443 [hep-th].

\bibitem{Wilhelm_Twisors_1}
L.~Koster, V.~Mitev, M.~Staudacher, M.~Wilhelm, \emph{	
All Tree-Level MHV Form Factors in N=4N=4 SYM from Twistor Space}, arXiv:1604.00012 [hep-th].

\bibitem{Wilhelm_Twisors_2}
L.~Koster, V.~Mitev, M.~Staudacher, M.~Wilhelm, \emph{Composite Operators in the Twistor Formulation of N=4N=4 SYM Theory}, arXiv:1603.04471 [hep-th].

\bibitem{LHC_1}
D.~Chicherin and E.~Sokatchev,
\emph{N=4 super-Yang-Mills in LHC superspace. Part I: Classical and quantum theory},
arXiv:1601.06803 [hep-th].

\bibitem{LHC_2}
D.~Chicherin and E.~Sokatchev,
\emph{N=4 super-Yang-Mills in LHC superspace. Part II: Non-chiral correlation functions of the stress-tensor multiplet},
arXiv:1601.06804 [hep-th].

\bibitem{LHC_3}
D.~Chicherin and E.~Sokatchev,
\emph{Composite operators and form factors in N=4 SYM},
arXiv:1605.01386 [hep-th].


\bibitem{BoFeng_BoundaryContributions}
Rijun~Huang, Qingjun~Jin, Bo~Feng, \emph{Form Factor and Boundary Contribution of Amplitude}, arXiv:1601.06612 [hep-th].

\bibitem{Wilhelm_Integrability_1}
M.~Wilhelm, \emph{Amplitudes, Form Factors and the Dilatation Operator in N=4SYM Theory}, JHEP 1502 (2015) 149, arXiv:1410.6309 [hep-th].

\bibitem{Wilhelm_Integrability_2}
D.~Nandan, C.~Sieg, M.~Wilhelm, Gang Yang, \emph{Cutting through form factors and cross sections of non-protected operators in N=4 SYM}, JHEP 1506 (2015) 156,
arXiv:1410.8485 [hep-th].

\bibitem{Wilhelm_Integrability_3}
M.~Wilhelm, \emph{Form factors and the dilatation operator in N=4 super Yang-Mills theory and its deformations}, arXiv:1603.01145 [hep-th].

\bibitem{Wilhelm_Integrability_4}
F.~Loebbert, D.~Nandan, C.~Sieg, M.~Wilhelm, Gang Yang, \emph{On-Shell Methods for the Two-Loop Dilatation Operator and Finite Remainders}, JHEP 1510 (2015) 012, arXiv:1504.06323 [hep-th].


\bibitem{FormFactorsSoftTheorems}
L.~V.~Bork and A.~I.~Onishchenko,
\emph{On soft theorems and form factors in $ \mathcal{N}=4 $ SYM theory},
JHEP {\bf 1512}, 030 (2015) , arXiv:1506.07551 [hep-th].


\bibitem{Wilhelm_Grassmannians_Integrability}
R.~Frassek, D.~Meidinger, D.~Nandan, M.~Wilhelm, \emph{On-shell diagrams, Gra\ss mannians and integrability for form factors}, JHEP 1601 (2016) 182, arXiv:1506.08192 [hep-th].



\bibitem{FF_ABJM_Young}
D.~Young, \emph{Form Factors of Chiral Primary Operators at Two Loops in ABJ(M)},
JHEP {\bf 1306} (2013) 049, arXiv:1305.2422 [hep-th].

\bibitem{FF_Sudakov_ABJM_Baianchi}
L.~Bianchi, M.~S.~Bianchi \emph{Non-planarity through unitarity in ABJM },
arXiv:1311.6464 [hep-th].

\bibitem{Penati_Santambrogio_ABJM_finite_N}
M.~S.~Bianchi, M.~Leoni, M.~Leoni, A.~Mauri,  S.~Penati and A.~Santambrogio,
\emph{ABJM amplitudes and WL at finite N}, JHEP {\bf 1309}, (2013) 114,
arXiv:1306.3243 [hep-th].

\bibitem{Brandhuber_ABJM_Sudakov2loops}
A.~Brandhuber, Omer~Gurdogan, D.~Korres, R.~Mooney, G.~Travaglini, \emph{Two-loop Sudakov Form Factor in ABJM}, JHEP 1311 (2013) 022, arXiv:1305.2421 [hep-th].

\bibitem{Henn_Different_Reg_FF}
J.~M.~Henn, S.~Moch, S.~G.~Naculich,
\emph{Form factors and scattering amplitudes in N=4 SYM in dimensional and massive regularizations},
JHEP {\bf 1112} (2011) 024, arXiv:1109.5057 [hep-th].

\bibitem{3loopSudakovN4SYM}
T.~Gehrmann, J.~M.~Henn and T.~Huber,
{\it The three-loop form factor in N=4 super Yang-Mills},
JHEP {\bf 1203}, 101 (2012)
[arXiv:1112.4524 [hep-th].

\bibitem{FF_Colour_Kinematic}
R.~H.~ Boels, B.~A.~Kniehl, O.~V.~Tarasov, Gang Yang,
\emph{Color-kinematic Duality for Form Factors},
JHEP {\bf 1302} (2013) 063, arXiv:1211.7028 [hep-th].

\bibitem{masters4loopSudakovN4SYM}
R.~Boels, B.~A.~Kniehl and G.~Yang,
{\it Master integrals for the four-loop Sudakov form factor},
arXiv:1508.03717 [hep-th].

\bibitem{Oluf_Tang_Engelund_Lagrangian_Insertion}
Oluf~Tang~Engelund, \emph{Lagrangian Insertion in the Light-Like Limit and the Super-Correlators/Super-Amplitudes Duality}, JHEP 1602 (2016) 030,  arXiv:1502.01934 [hep-th].

\bibitem{Drummond_Grassmannians_Tduality}
  J.~M.~Drummond and L.~Ferro,
  \emph{Yangians, Grassmannians and T-duality},
  JHEP {\bf 1007} (2010) 027,
  [arXiv:1001.3348 [hep-th].

\bibitem{Drummond_Yangian_origin_Grassmannian_integral}
  J.~M.~Drummond and L.~Ferro,
  \emph{The Yangian origin of the Grassmannian integral},
  JHEP {\bf 1012} (2010) 010,
  [arXiv:1002.4622 [hep-th].

\bibitem{TotalPositivityGrassmanniansNetworks}
A.~Postnikov,
{\it Total positivity, Grassmannians, and networks},
math/0609764 [math.CO].

\bibitem{BipartiteFieldTheories}
S.~Franco,
{\it Bipartite Field Theories: from D-Brane Probes to Scattering Amplitudes},
JHEP {\bf 1211}, 141 (2012),
arXiv:1207.0807 [hep-th].

\bibitem{HarmonicRmatrices}
L.~Ferro, T.~Lukowski, C.~Meneghelli, J.~Plefka and M.~Staudacher,
{\it Harmonic R-matrices for Scattering Amplitudes and Spectral Regularization},
Phys.\ Rev.\ Lett.\  {\bf 110}, no. 12, 121602 (2013),
arXiv:1212.0850 [hep-th].


\bibitem{Broedel_DictionaryRoperatorsOnshellGraphsYangianAlgebras}
J.~Broedel, M.~de Leeuw and M.~Rosso,
{\it A dictionary between R-operators, on-shell graphs and Yangian algebras},
JHEP {\bf 1406}, 170 (2014)
[arXiv:1403.3670 [hep-th].

\bibitem{SoftTheoremsGrassmannian}
J.~Rao,
\emph{Soft theorem of $ \mathcal{N} $ = 4 SYM in Grassmannian formulation},
JHEP {\bf 1502}, 087 (2015), arXiv:1410.5047 [hep-th].

\bibitem{N=4_Harmonic_SS}
G.~G.~Hartwell, P.~S.~Howe, \emph{(N, p, q) harmonic superspace},
Int.\ J.\ Mod.\ Phys.\ A {\bf 10} (1995) 3901-3920, hep-th/9412147.

\bibitem{SuperCor1}
B.~Eden, P.~Heslop, G.~P.~Korchemsky, E.~Sokatchev,
\emph{The super-correlator/super-amplitude duality: Part I},
arXiv:1103.3714v1 [hep-th].

\bibitem{SM}
  D.~Maitre and P.~Mastrolia,
  \emph{S @ M, a Mathematica Implementation of the Spinor-Helicity Formalism},
  Comput.\ Phys.\ Commun.\  {\bf 179} (2008) 501,
  arXiv:0710.5559 [hep-ph].

\bibitem{Grassmanians-N4SYM-ABJM}
H.~Elvang, Y.~t.~Huang, C.~Keeler, T.~Lam, T.~M.~Olson, S.~B.~Roland and D.~E.~Speyer,
\emph{Grassmannians for scattering amplitudes in 4d $\mathcal{N}=4$ SYM and 3d ABJM},
JHEP {\bf 1412}, 181 (2014), arXiv:1410.0621 [hep-th].

%
%
%
%
%
%
%
%
%
%
%
%
%
%
%
%
%
%
%
%
%
%
%
%
%
%
%
%
%
%
%
%
%
%
%
%
%
%
%
%
%
%
%
%
%
%
%
%
%
%
%
%
%
%
%
%
\end{thebibliography}
\end{document}